\documentclass[10pt,amssymb,superscriptaddress,showkeys,twocolumn,prl]{revtex4-2}

\usepackage{graphicx}
\usepackage{dcolumn}
\usepackage{bm}
\usepackage{amsmath,amssymb}
\usepackage{titlesec}
\usepackage{lineno}
\usepackage{hyperref}

\makeatletter
\renewcommand{\@caption@fignum@sep}{\space} 
\makeatother
\titleformat{\section}[block]{\normalfont\bfseries\large}{\thesection}{1em}{}
\titlespacing*{\section}{0pt}{\baselineskip}{0.5\baselineskip}

\titleformat{\subsection}[block]{\normalfont\bfseries\small}{\thesubsection}{1em}{}
\titlespacing*{\subsection}{0pt}{\baselineskip}{0.5\baselineskip}

\renewcommand{\figurename}{\textbf{Fig.}}  

\hypersetup{
 colorlinks=true,
 linkcolor=blue,
 filecolor=blue,  
 urlcolor=blue,
 citecolor=blue,
}

\begin{document}

\title{Deep-Learning-Empowered Programmable Topolectrical Circuits}
\author{Hao Jia}
\thanks{These authors contributed equally}
\affiliation{School of Physical Science and Technology, Lanzhou University, Lanzhou 730000, China.}
\author{Shanglin Yang}
\thanks{These authors contributed equally}
\affiliation{School of Optoelectronic Engineering, Xidian University, Xi'an 710071, China}
\author{Jiajun He}
\affiliation{School of Physical Science and Technology, Lanzhou University, Lanzhou 730000, China.}
\author{Shuo Liu}
\affiliation{State Key Laboratory of Millimeter Waves, Southeast University, Nanjing 210096, China.}
\author{Haoxiang Chen}
\affiliation{International School of Microelectronics, Dongguan University of Technology, Dongguan 523808, Guangdong, China.}
\author{Ce Shang}
\email{shangce@aircas.ac.cn}
\affiliation{Aerospace Information Research Institute, Chinese Academy of Sciences, Beijing 100094, China}
\author{Shaojie Ma}
\affiliation{Shanghai Engineering Research Centre of Ultra Precision Optical Manufacturing, Department of Optical Science and Engineering, School of Information Science and Technology, Fudan University, Shanghai 200433, China}
\author{Peng Han}
\affiliation{School of Computer Science and Engineering, University of Electronic Science and Technology of China, Chengdu 611731, China}
\author{Ching Hua Lee}
\affiliation{Department of Physics, National University of Singapore, Singapore 117551, Republic of Singapore}
\author{Zhen Gao}
\affiliation{Department of Electronic and Electrical Engineering, Southern University of Science
and Technology, Shenzhen 518055, China.}
\author{Yun Lai}
\email{laiyun@nju.edu.cn}
\affiliation{National Laboratory of Solid State Microstructures, School of Physics, and Collaborative Innovation Center of Advanced Microstructures, Nanjing University, Nanjing 210093, China}
\author{Tie Jun Cui}
\email{tjcui@seu.edu.cn}
\affiliation{State Key Laboratory of Millimeter Waves, Southeast University, Nanjing 210096, China.}
\date{\today}

\begin{abstract}
Topolectrical circuits provide a versatile platform for exploring and simulating modern physical models. However, existing approaches suffer from incomplete programmability and ineffective feature prediction and control mechanisms, hindering the investigation of physical phenomena on an integrated platform and limiting their translation into practical applications. Here, we present a deep-learning-empowered programmable topolectrical circuits (DLPTCs) platform for physical modeling and analysis. By integrating fully independent, continuous tuning of both on-site and off-site terms of the lattice Hamiltonian, physics-graph-informed inverse state design, and immediate hardware verification, our system bridges the gap between theoretical modeling and practical realization. Through flexible control and adiabatic path engineering, we experimentally observe the boundary states without global symmetry in higher-order topological systems, their adiabatic phase transitions, and the flat-band-like characteristic corresponding to Landau levels in the circuit. Incorporating a physics‑graph‑informed mechanism with a generative AI model for physics exploration, we realize arbitrary, position-controllable on-board Anderson localization, surpassing conventional random localization. Utilizing this unique capability with high‑fidelity hardware implementation, we further demonstrate a compelling cryptographic application: hash-based probabilistic information encryption by leveraging Anderson localization with extensive disorder configurations, enabling secure delivery of full ASCII messages. 
\end{abstract}

\maketitle
\section*{Main}
The eternal endeavor of human beings is to search for universal tools that facilitate the comprehension and elucidation of our world through a broad spectrum of disciplines, including mathematical equations, scientific laws, theoretical frameworks, and experimental approaches. With such tools, scientists are empowered to forecast outcomes, test hypotheses, and continually expand our grasp of the universe. Quantum simulators \cite{feynman2018simulating}, a concept pioneered by Richard Feynman in 1982, propose Hamiltonian simulation as a solution for emulating physical complexity in a programmable manner. Quantum simulators have been realized on several experimental platforms, including systems of ultra-cold quantum gases \cite{cooper2019topological}, trapped ions \cite{Shao2024}, photonic systems \cite{Bogaerts2020, RevModPhys.91.015006, On2024, dai2024programmable, PhysRevLett.129.140502, koh2024realization}, plasmonic systems \cite{You2021}, and superconducting circuits \cite{PhysRevLett.127.180501}, which are particularly adept at simulating topological physics with a high degree of reconfigurability. 

\begin{figure*}[t!]
    \centering
    \includegraphics[width=1\linewidth]{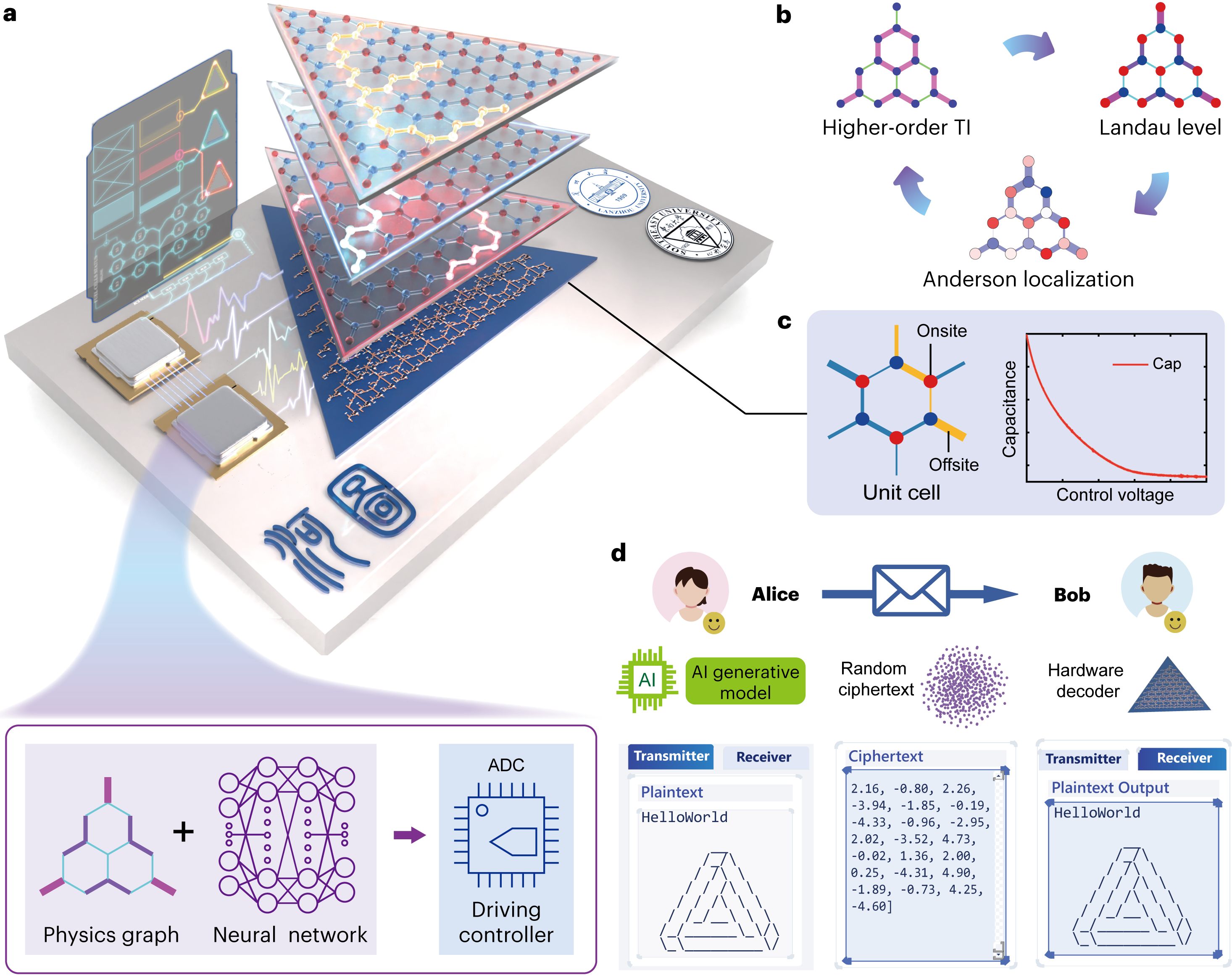}
    \caption{\textbf{The conception of DLPTCs ‘HeTu’.} a, The schematic of DLPTC and integration with different components. Physics-graph information is embedded into the deep learning framework for physics phenomenon prediction and generation. The generated parameters are loaded into the driving controller and used to drive the DLPTC. b, By controlling the onsite and offsite terms, this setup facilitates various complex systems, including higher-order topological lattices, flat-band systems for Landau levels, and Anderson localizations, with further potential for exploration beyond these areas. c, The on-site and off-site terms, corresponding to the vortex and edge values of the graph, are fully programmable through voltage control. d, With the DLPTC, we demonstrate an ASCII-based message encryption system for practical application.}
    \label{fig1}
\end{figure*}

To date, a significant challenge of programmable Hamiltonian simulators is the observation of multiple complex phenomena, such as topological phase transitions \cite{RevModPhys.82.1959, RevModPhys.83.1057, Benalcazar61}, Landau levels \cite{rechtsman2013strain, PhysRevLett.118.194301,barsukova2024direct}, and Anderson localization \cite{Segev2013, Yang2024}. So far, achieving precise control over both the on-site potential and the off-site hopping strength across a multitude of samples remains a daunting task. In response to this limitation, we propose fully programmable topolectrical circuits (PTCs) that allow for independent control over the on-site and off-site terms within separate degrees of freedom. By granting full control over the circuit Laplacian, our platform enables the simulation of a diverse range of Hamiltonian systems, thereby facilitating a more straightforward investigation of a wide range of physical phenomena and mechanisms \cite{Lee2018, Hadad2018, Wu2022}. Building on this capability, we integrate an artificial intelligence (AI) module into the circuitry to create deep learning-empowered programmable topolectrical circuits (DLPTCs). This integration not only provides effective feature prediction mechanisms and promotes the investigation of physics, but also bridges the gap between emerging physical phenomena and their practical applications.

Recent advances in deep learning have proven invaluable for feature extraction and prediction in complex physical systems \cite{carleo2019machine, raissi2020hidden, karniadakis2021physics}. However, traditional neural network architectures, including fully connected neural networks (FCNNs)\cite{lecun2015deep}, convolution neural networks (CNNs) \cite{gu2018recent}, along with their derivatives \cite{he2016deep}, often struggle to achieve satisfactory performance due to the inherent high disorder and intricacies of these systems. To overcome this bottleneck, we have proposed physics-graph-informed convolution neural networks (PGI-CNNs) and physics-graph-informed generative model. By embedding the tight-binding information, fundamental physical laws \cite{karniadakis2021physics, Shang2022}, and domain knowledge of circuit topology \cite{shang2024observation} as intrinsic network priors, these models achieve superior predictive and generative performance in capturing the nuances of disordered systems, particularly in reliably forecasting and generating Anderson localization characteristics.

Through flexible control and adiabatic path engineering, we experimentally observe the boundary states without global symmetry in higher-order topological systems, their adiabatic phase transitions, and the flat-band-like characteristic corresponding to Landau levels in the circuit. Incorporating a PGI mechanism with a generative AI model, we realize arbitrary, position-controllable on-board Anderson localization, surpassing conventional random localization. Among them, the PGI-CNN model is utilized for the analysis and understanding of the Anderson system to predict the localization characteristic. A generative PGI-diffusion and PGI-conditional variational autoencoder model has been developed for the generation of arbitrary localization, enabling a unique information encryption framework. We further developed a universal ASCII message encryption system that harnesses the unique capability with high‑fidelity hardware implementation to transmit information between two universities.  These achievements underscore the far-reaching potential of programmable Hamiltonian simulators enhanced by deep learning, offering a versatile platform for both exploring physics and enabling innovative applications in interdisciplinary research.

\section*{Results}
The prototype of the DLPTCs platform is hereby named ‘HeTu’ (‘River Chart’ in English, which originates from an ancient Chinese civilization legend, meaning a diagram consists of trigrams, points, and connections, and can be used to explore unknown things and reveal the mechanisms).
By allowing researchers to dynamically design, simulate, and control the Hamiltonian with greater precision and artificial intelligence, this platform could provide groundbreaking insights into phases of matter and pave the way for the invention of innovative applications. The base hardware structure is a honeycomb-like lattice, characterized by off-site hopping $E$ and on-site potential $V$ strengths in a $(N+1)^2$-site distributed in triangular geometry, as the conception Figure \ref{fig1}a shows. The sites are designated as $A_{i j k}$ and $B_{i j k} $ for the sublattices $\mathrm{A}$ $(\xi=-1)$ and $\mathrm{B}$ $(\xi=1)$ (shown in red and blue circles), with $i, j$, and $k$ being the indices in the directions $\mathbf{e}_1=(-\sqrt{3} / 2,-1 / 2)$, $\mathbf{e}_2=(\sqrt{3} / 2,-1 / 2)$, and $\mathbf{e}_3=(0,1)$, following $i+j+k+(\xi+1) / 2=N$, respectively.
 
The couplings of such lattices (hopping) are described by the tight-binding model (TBM):
\begin{equation}\label{eq:1}
H_{\rm {hopping}}=\sum_{{\mathbf{r}}= (i, j, k)}\sum_{d= 1,2,3}E_{\mathbf{e}_d}(\mathbf{r}) b_{\mathbf{r}-\mathbf{e}_d}^{\dagger}a_{\mathbf{r}}+\text { h.c. },
\end{equation}
and the potential is described by
\begin{equation} \label{eq:2}
H_{\rm {potential}}=\sum_{{\mathbf{r}}= (i, j, k)}\left[V_a(\mathbf{r}) a_{\mathbf{r}}^{\dagger}a_{\mathbf{r}}+  V_b(\mathbf{r}) b_{\mathbf{r}}^{\dagger}b_{\mathbf{r}}\right],
\end{equation}
where $a_{i j k}$ and $b_{i j k}$ are the annihilation operators of $A_{i j k}$ and $B_{i j k}$ sites, respectively. According to graph theory, the network can be converted into a weighted graph $G=[V, E]$, where $V, E$ represent the set of on-site and off-site terms. At the electric circuits level, the theory ensures that one can fully map the tight-binding Hamiltonian in condensed matter physics to circuit Laplacians (admittance matrices) (Supplementary Information S1.A). We employ voltage-controlled variable capacitors to implement both the on-site and off-site (coupling) terms, whose values can be continuously tuned over the positive real range. Leveraging this capability, we develop a hardware–software synergy framework that supports an AI-fusion circuit platform for the implementation of physical systems, as well as for interdisciplinary research and application exploration (Supplementary Information S1.B\&C). Furthermore, by engineering the hopping units and integrating active components, we can extend the tunability to encompass the full ranges of negative- and complex-valued parameters (Supplementary Information, S1.D).

\section*{Higher-order topological insulator without global symmetry}
By engineering both intracell and intercell hoppings, the exploration of HOTIs becomes feasible. HOTIs are a new class of topological insulators with dimensions $d>1$ and have attracted considerable interest as a novel topological phase of matter \cite{Benalcazar61}. A hallmark of higher-order topological phases is the presence of topological boundary states determined by the symmetry of the unit cell \cite{Bradlyn2017, zou2024experimental}. In contrast, we present an HOTI without global symmetry, which contains a structure with complete and incomplete unit cells, and show that the higher-order topologically protected mechanism exists in it.

\begin{figure*}[htbp!]
    \centering
    \includegraphics[width=1\linewidth]{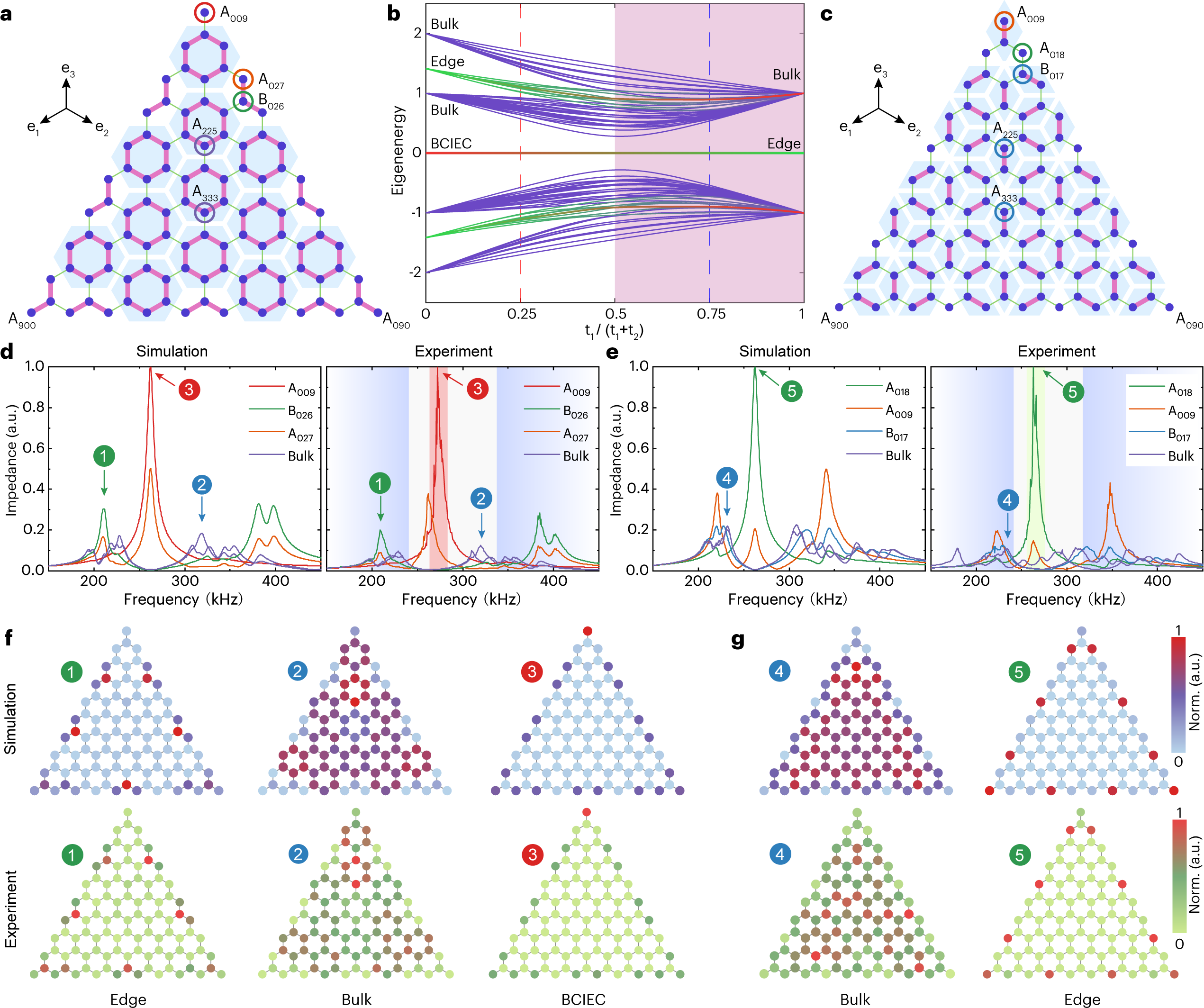}
    \caption{\textbf{Higher-order topological insulator without global symmetry.}  \textbf{a}, Schematic of the lattice model when $t_1<t_2$. \textbf{b}, The energy spectrum for a function of $t_1/(t_1+t_2)$.  \textbf{c}, Schematic of the lattice model when $t_1>t_2$. \textbf{d}, The calculated and measured impedance spectra on characteristic points of model \textbf{a}. \textbf{e}, The calculated and measured impedance spectra on characteristic points of model \textbf{b}. \textbf{f}, The impedance distribution of edge, bulk, and BCIEC modes in model \textbf{a}. \textbf{g}, The impedance distribution of bulk and edge modes in model \textbf{b}.}
    \label{fig2:higher-oeder}
\end{figure*}

\begin{figure*}[t]
    \centering
    \includegraphics[width=1\linewidth]{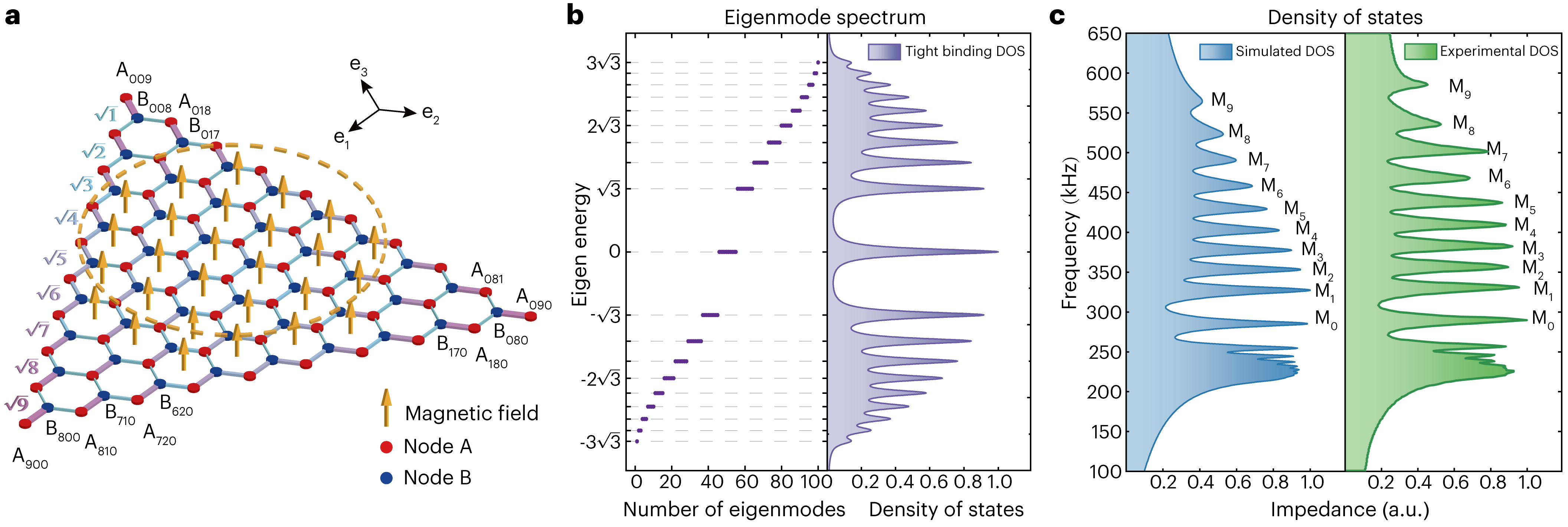}
    \caption{\textbf{Two-dimensional Fock-state lattice with an effective pseudomagnetic field and Landau levels.} \textbf{a}, Schematic of the flat-band lattice. \textbf{b}, The eigenenergy spectrum and corresponding DOS calculated by TBM. \textbf{c}, Circuit simulation and experiment result of DOS.}
    \label{fig3:Landau}
\end{figure*}

Here, the HOTI without global symmetry is implemented on the PTCs by making the Kekul\'{e}-type and Anti-Kekul\'{e}-type \cite{noh2018topological} spatial modulation of hopping terms and unconventional boundary truncations. The equation \ref{eq:1} is programmed as
\begin{equation} 
\begin{aligned} 
    E_{\mathbf{e}_1} &= t_1 + (-1)^{\text{mod}(i-k-1, 3)} t_2, \\ 
    E_{\mathbf{e}_2} &= t_1 + (-1)^{\text{mod}(j-i-2, 3)} t_2, \\ 
    E_{\mathbf{e}_3} &= t_1 + (-1)^{\text{mod}(k-i, 3)} t_2,
\end{aligned} 
\end{equation} 
with all terms in equation (\ref{eq:2}) kept the same. The model is implemented on a circuit lattice with a scale $N=9$, as shown in Figure \ref{fig2:higher-oeder}a ($t_1$ \textless $t_2$) and Figure \ref{fig2:higher-oeder}c ($t_1$ \textgreater $t_2$). The detailed theory and analysis are given in Supplementary Information S2.A and S2.B. The topological phase diagram is calculated as a function of the parameter $\gamma= t_1/(t_1+t_2)\in (0,1)$, shown in Figure \ref{fig2:higher-oeder}b (Supplementary Information S2.C). Unlike HOTIs with global symmetry, where the topological nontrivial phase only occurs at $t_2 > t_1$, in HOTIs without global symmetry, boundary states are found both at $t_2 < t_1$ and $t_2 > t_1$. We focus on two cases: I) $t_2=3t_1$ $(\gamma=0.25)$  and II) $t_1=3t_2$ $(\gamma=0.75)$ as indicated by the red and blue dashed lines in Figure \ref{fig2:higher-oeder}b. In case I, except for degenerate bulk and edge states, there exist novel bound corner states in the edge continuum (BCIEC), indicating that the corner states are observed to reside within the edge band. With the increase of $\gamma$, all BCIECs transition into edge states in case II. There is still a transition point $t_2 = t_1$, which distinguishes between case I and case II phases separated by the transition point $\gamma=0.5$. To thoroughly investigate the phase transition process, we varied the parameter $\gamma$ from 0.25 to 0.75 adiabatically, and the results are given in Extended Data Fig. \ref{ext_HOTI}.

The corresponding model is mapped to the PTCs platform for further study (Supplementary Information S2.D). The impedance spectra are simulated and measured, as shown in Figure \ref{fig2:higher-oeder}d and \ref{fig2:higher-oeder}e. The peak and normalized intensities show good agreement, with minor shape deviations and peak extension in experimental data attributed to inevitable fluctuation. In case I, the peak at point $A_{009}$ is marked as \textcircled{3}. The impedance distribution over all sites at this frequency illustrates the excited mode, as shown in Figure \ref{fig2:higher-oeder}f, in which the BCIEC mode is excited. The peak at point $A_{018}$ is marked as \textcircled{5}, which can also be verified from the mode distribution in Figure \ref{fig2:higher-oeder}g. Similarly, the edge states in case I and the bulk and edge states in case II are demonstrated. By adiabatically tuning the off-site capacitance, we demonstrate the phase transition process. Such a demonstration reveals that our PTC is capable of freely tuning the hopping strength to dynamically control the topological phase and promote the discovery of new topological states such as BCIEC.

\section*{Flat bands and Landau levels on PTC platform}
Beyond its application in the HOTIs model, our PTC's ability to fully control off-site terms allows us to realize more complex physical models other than topological phenomena. One such scenario is presented by the Flat bands model, which requires gradient tuning and precise engineering of off-site terms. Flat bands in matter could lead to superconductivity \cite{cao2018unconventional},  nontrivial topology \cite{cao2018correlated, PhysRevLett.121.237401}, and localization \cite{wang2020localization}, garnering growing interest and potential applications. In particular, by engineering the hopping strengths between lattice sites, it is possible to realize all-band-flat (ABF) lattices \cite{Cai2020, yang2024realization, deng2022observing}. However, due to limitations in achievable coupling strengths, efforts to aggregate all eigenstates into flat bands in experiments are relatively scarce. Nevertheless, with the PTC platform, we can precisely control the coupling terms to construct an ABF honeycomb lattice in Figure \ref{fig3:Landau}a. The equation (\ref{eq:1}) is programmed as
\begin{equation} 
\begin{aligned} 
    E_{\mathbf{e}_1},  E_{\mathbf{e}_2},  E_{\mathbf{e}_3}  = \sqrt{i} t_0, \sqrt{j} t_0, \sqrt{k} t_0
\end{aligned} 
\end{equation} 
with all terms in equation (\ref{eq:2}) kept the same. 

The eigenenergies can be solved analytically as 
\begin{equation}
    \epsilon_m= \pm \sqrt{3 m} t_0
    \label{eq:landau_energy_level}
\end{equation}
with $m=0,1,2, \ldots, N$ and all eigenstates are organized into $2N+1$ flat bands. Figure \ref{fig3:Landau}b illustrates the calculated eigenenergy spectrum and density of states (DOS) through TBM. Corresponding to the center of each peak in the DOS, 19 flat bands are observed. The ten peaks for energy $E_0$ to $E_9$ can be observed. For circuit implementation (Supplementary Information S3.A\&B). The impedance spectra of each site are proportional to the local density of states (LDOS). The impedance curve for all sites is simulated with parasite parameters. With minor parasite parameters, all 19 impedance bands can be observed, corresponding to the peaks of TBM DOS. Due to circuit frequency dispersion, the lower half of the bands below the central frequency is compressed, while the upper half is stretched. Consequently, the nine peaks indicating $M_1$ to $M_9$ are widely separated, while the other peaks indicating $M_{-1}$ to $M_{-9}$ are closely packed(Extended Data Fig. \ref{ext_Landau}). The results of the simulation and experiments with actual parasite parameters are shown in Figure \ref{fig3:Landau}c and match very well. Although the impedance bands can be observed, further broadening of the impedance bands induces partial overlap of peaks at lower frequencies. 

We further measure and visualize the modes within flat bands. The excitation frequencies for $M_0$ to $M_9$ are selected based on DOS. The spatial distribution for each Landau level is calculated and measured. The results from TBM, circuit simulation, and experimental measurements are in strong agreement. Each mode tends to occupy specific locations and exhibits unique characteristics (Supplementary Information S3.C). This validates our platform as an analog for flat-band models. Although the circuit system lacks Coulomb interactions that give rise to interaction-driven many-body correlated phases in electronic materials, Hamiltonian mapping still allows the circuit Laplacian to be engineered to reproduce the eigen‑spectrum of target quantum models, including those with effective interactions \cite{Olekhno2020Topolectrical,zhou2023observation}.

\begin{figure*}[htbp!]
    \centering
    \includegraphics[width=\linewidth]{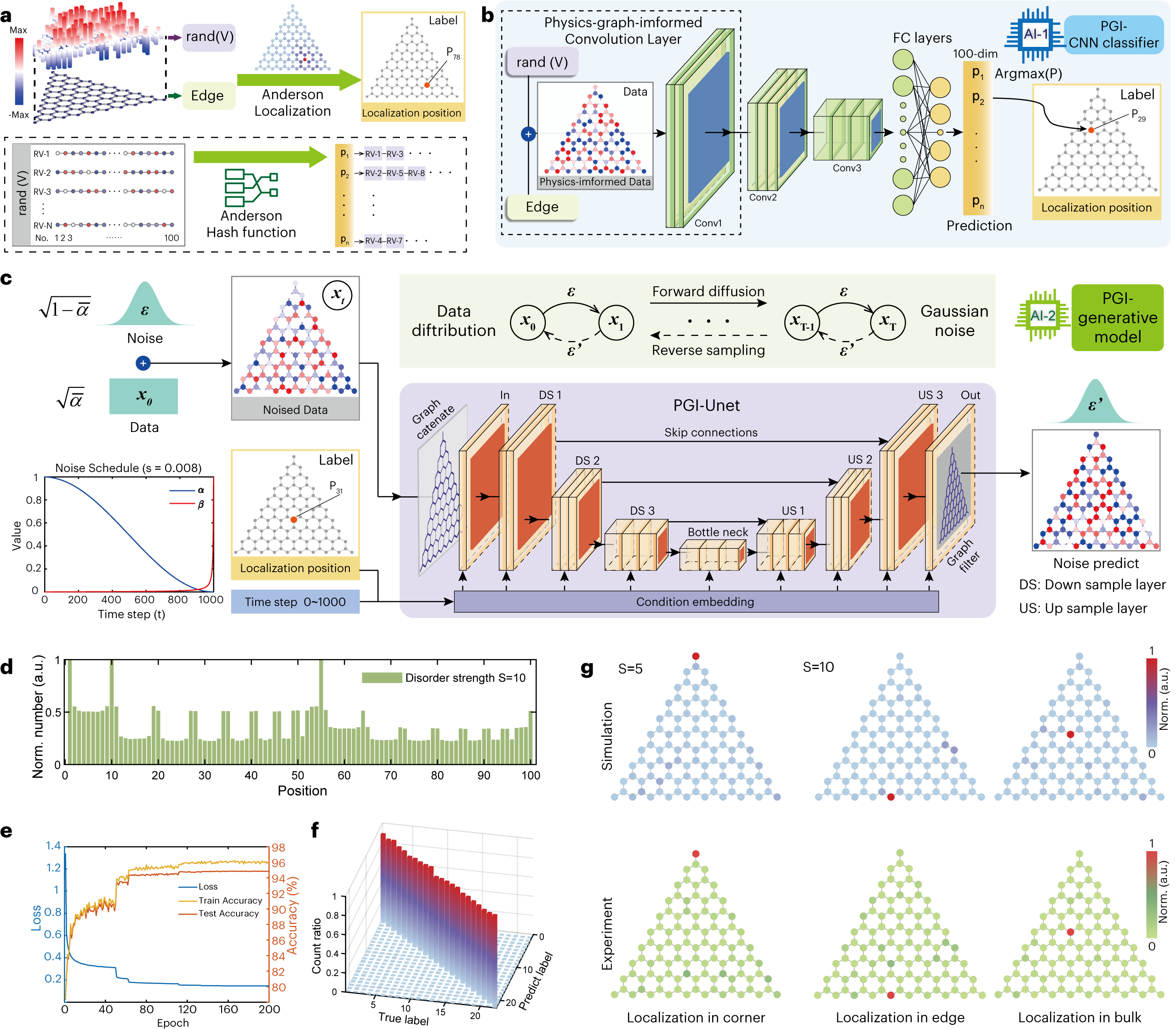}
    \caption{ \textbf{Deep-learning empowered controllable Anderson Localization generation on DLPTCs.} \textbf{a},  Anderson localization and the corresponding Hash characteristic. \textbf{b},  PGI-CNN classifier for localization position prediction. \textbf{c},  PGI-generative model for Anderson Localization generation with arbitrary localized position. The overall model employs a U-Net structure with residual blocks and conditional mechanisms as its backbone. In the forward diffusion, noise is progressively added to the PGI-structured training data sample until it resembles pure Gaussian noise. In the reverse sampling, the model iteratively removes the noise, generating the data to achieve conditional position Anderson localization. \textbf{d}, Normalized localization probability on each site with $S=10$ using PGI-classfier. \textbf{e}, The training loss and prediction accuracy for PGI-CNN, which shows that the accuracy on arbitrary test data can achieve higher than 95\%. \textbf{f}, The confusion matrix of generated data using a generative model, the diagonal line, meaning correct data, over 94\%, which can ensure correct data encryption with an error correction checking mechanism. \textbf{g}, The simulated and experimental mode distribution, the localization position matches well and can be on an arbitrary position for $S=10$.  }
    \label{fig4:anderson_phys}
\end{figure*}

\section*{Deep Learning-Empowered Generation of Controllable Anderson Localization}
Maximizing the reconfigurability of PTCs involves the arbitrary control of the on-site and off-site hopping terms, which allows us to observe and control the Anderson localization phenomena, especially the realization of freely tailoring the strong localization in limited systems with higher dimensions. The concept of Anderson localization, first introduced by Anderson in 1958, has been pivotal in understanding the behavior of wave functions in disordered systems \cite{physRev.109.1492}. It illustrates that disorder can lead to the localization of any wave function \cite{kondov2011three}. The 'dimension curse' of Anderson localization makes localization more difficult due to the dissipation of energy along multiple dimensions. It is especially challenging to discern Anderson localization in small, confined regions. Generally, statistical methods must be contemplated with a diverse array of samples to gain a more comprehensive understanding of this phenomenon \cite{yu2021engineered}. This challenge underscores the need for more advanced theoretical frameworks and experimental techniques to reduce the demands on the system and more easily and directly control and observe this phenomenon. 

Considering on-site disorder in Anderson's theory, the equation (\ref{eq:1}) is programmed as
\begin{equation} 
\begin{aligned} 
    V_{\mathbf{a}},  V_{\mathbf{b}} =  S \beta_{\mathbf{r}_a} , S \beta_{\mathbf{r}_b},
\end{aligned} 
\end{equation} 
The on-site terms $\{\beta_{\mathbf{r}_a},\beta_{\mathbf{r}_b}\} \in[-1/2, 1/2]$ are sampled from the uniform random distribution, described by $G=[{\rm{rand}}(V), E]$, and $S$ is a coefficient indicating disorder strength.

Initially, the hashing characteristic of Anderson localization—a phenomenon where multiple types of disorder configurations correspond to a single type of localization positions—is illustrated in Figure \ref{fig4:anderson_phys}a. To elucidate this, we generate various combinations of ${\rm{rand}}(V)$ and find that the wave function can be made highly concentrated at a specific site, as indicated in the yellow frame. Hence, the localization position is a unique label (identifier) for these disorder configurations. We have developed a physics-informed data screening algorithm to create dataset for AI model (details of the dataset can be found in Supplementary Information S4.A \& S4.B).

One of the key characteristics representing the localization phenomenon is the maximum position of localization. Deep learning methods can be introduced for position prediction instead of solving the system's wave function, particularly for arbitrary disorder data that can induce highly concentrated localization. However, due to the highly disordered system characteristic and complexity of connections, commonly used FCNNs \cite{lecun2015deep}, CNNs\cite{gu2018recent}, and ResNets \cite{he2016deep}, can hardly achieve satisfactory performance on the dataset. To solve this issue, physics-graph information, which is pivotal for such a system, should be considered in network construction. Moreover, tight-binding rules should also be embedded as a physics constraint. Therefore, we propose a PGI mechanism, which embeds tight-binding rules and graph information into the network (Extended Data Fig. \ref{ext_AI}). Using such a mechanism, we construct a PGI-CNN for localization prediction. As Figure \ref{fig4:anderson_phys}b shows, the disorder parameters ${\rm{rand}}(V)$ is combined with the graph edge to obtain physics-informed data. Then the first layer of the network employs a physics-graph-informed convolution layer while without a pooling layer, to capture tight-binding features that indicate the coupling between nearest neighbor sites. Subsequent layers include additional convolution layers with pooling for dimension reduction and feature extraction, followed by fully connected layers for output prediction. Dataset enhancement based on rotational symmetry of geometry (Supplementary information S4.B) is introduced during the training process. After that, the predicted precision in the parameters of the arbitrary disorder that can induce strong localization is achieved at over 95\% (Figure \ref{fig4:anderson_phys} e), significantly higher compared to networks that do not utilize the PGI mechanism. A detailed discussion on PGI-CNNs design, the training process, and the comparison is given in Supplementary Information S4.C \& S4.D. 

Using PGI-CNNs as a predictor, we conduct a statistical analysis of the distribution of localized positions across the disorder amplitude range of [-5, 5] in Figure \ref{fig4:anderson_phys}d. The results indicate that the clustered points exhibit positional selectivity. The three corner sites exhibit the highest probability, indicating that energy tends to concentrate more readily at these locations. The sites along the three edges have the next highest probabilities, while the sites within the bulk have the lowest probabilities. As the disorder strength is further increased, the probability of localization at bulk positions also increases. More results are provided in Supplementary Information S4.E.

Beyond prediction, we have developed a PGI-generative model to enable efficient, arbitrary control of the localization phenomenon. As illustrated in Figure \ref{fig4:anderson_phys}c, this network combines the advantages of PGI with state-of-the-art diffusion models \cite{ho2020denoising}. The spatial structure of the topological graph and the tight-binding rules are embedded by the PGI-mechanism to provide intrinsic physical information. The overall model employs a U-Net structure with residual blocks and conditional mechanisms as its backbone. In the forward diffusion process(training), the model takes a precise data sample $x_0=(z_{10}, ...z_{n0}) $ from the uniform dataset and progressively adds Gaussian noise over a sequence of steps, producing a noise sample $x_t$. The sample is increasingly corrupted, eventually converging towards a normal distribution. In the reverse sampling process, the model iteratively predicts and removes the noise step by step, thereby transforming $x_t$ back into $x_0$. This denoising process ultimately produces a realistic  ${\rm{rand}}(V)$ from the input noise sample and label. Further details are provided in Extended Data Fig. \ref{ext_AI}d and Supplementary Information S4.F). Figure \ref{fig4:anderson_phys}f depicts the confusion matrix, which evaluates the accuracy of the generated data for each localization position, demonstrating a high level of consistency with the labels. The average accuracy for all labels exceeds 94\%. With such high accuracy, it is easy for message coding with some error-correcting mechanisms deployed \cite{santini2020analysis}. We also implement the physics information embedding on classical generative network conditional Variational Autoencoder (cVAE) \cite{sohn2015learning}, which also shows some advances, and the accuracy can achieve 81\%. (Supplementary Information S4.H). 

To further validate the accuracy and applicability of our method, we have loaded the generated data into the PTC platform. Representative simulation and experimental localization results are depicted \ref{fig4:anderson_phys}g,  where the localization positions and intensities show an excellent agreement across all cases. Additional data and discussion are provided in Supplementary Information S4.I. Our DLPTC, implemented with a PGI-CNN classifier and a PGI-generative model, offers a powerful toolkit for analyzing and engineering Anderson localization with flexible design capabilities.

\section*{Programmable Anderson Localization and Hash-based probabilistic Information Encryption}

\begin{figure*}[htbp!]
    \centering
    \includegraphics[width=\linewidth]{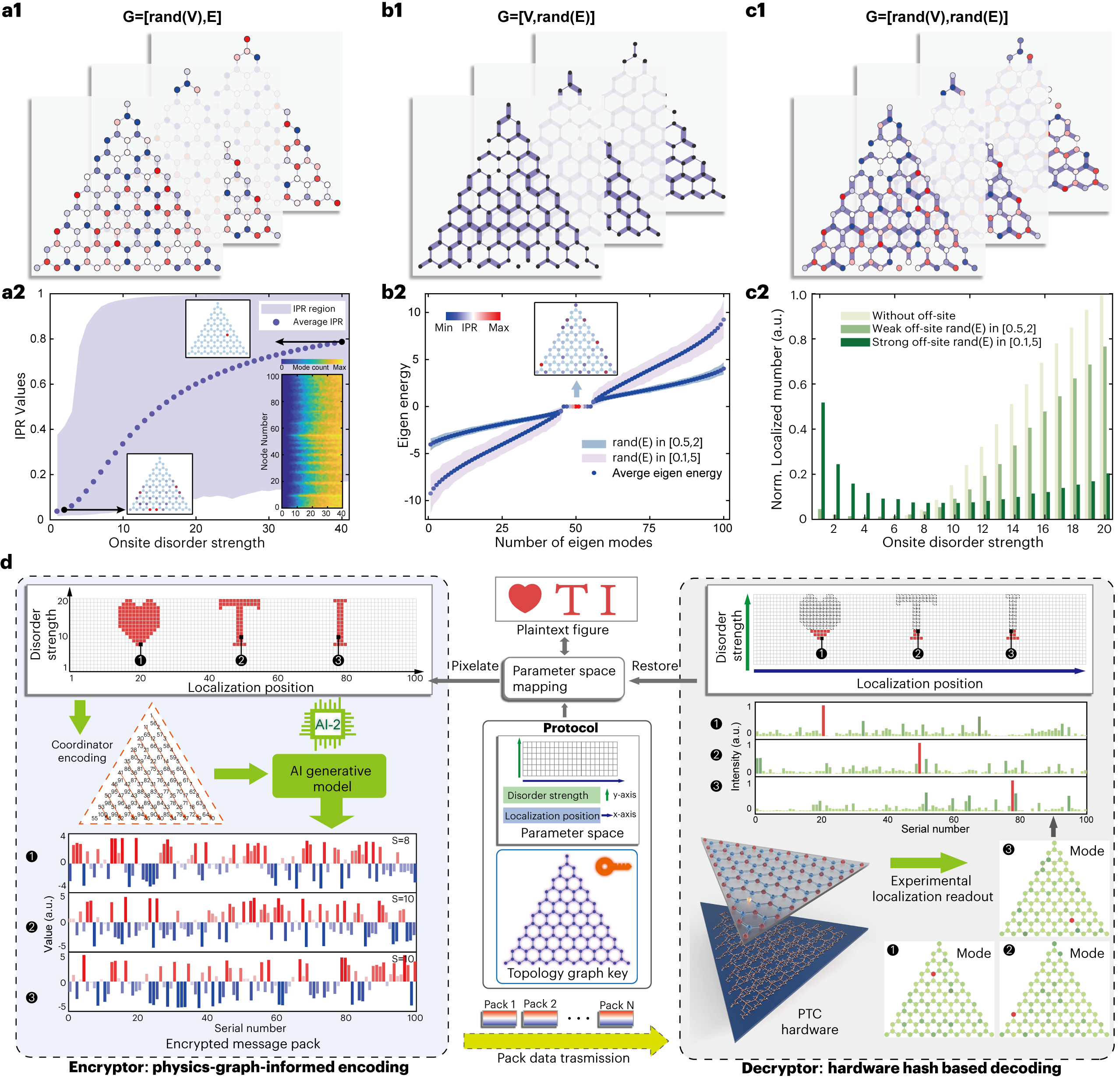}
    \caption{\textbf{Three types of Anderson localization phenomena and Hash-based probabilistic Encryption Application using Programmable Anderson Localization.} \textbf{a1}, Only on-site term disorder and localization performance analysis \textbf{a2}. \textbf{b1}, Only off-site term disorder and the eigen-mode spectra \textbf{b2}, which shows multiple-points localization characteristics. \textbf{c1}, Both on-site and off-site term disorders and their synergistic and antagonistic contribution to the localization performance \textbf{c2}. \textbf{d}, The encryption pipeline and experimental demonstration of encrypted picture transmission with meaning 'love topological insulator'.}
    \label{fig5:anderson_cipher}
\end{figure*}

Owing to the combination of powerful programmable hardware and algorithm-enhanced software, we are not only able to freely tailor conventional physical phenomena but also to discover new physics. When the on-site and off-site terms can all be programmed as
\begin{equation} 
\begin{aligned} 
    E_{\mathbf{e}_1},  E_{\mathbf{e}_2},  E_{\mathbf{e}_3}  &= \beta_{\mathbf{e}_1} , \beta_{\mathbf{e}_2}, \beta_{\mathbf{e}_3}, \\
    V_{\mathbf{a}},  V_{\mathbf{b}} &= S \beta_{\mathbf{r}_a} , S \beta_{\mathbf{r}_b}.
\end{aligned} 
\end{equation} 
The phenomenon of Anderson localization can be further categorized into three distinct types, as illustrated in Figure \ref{fig5:anderson_cipher}a1, b1, c1, characterized by only on-site term disorder $G=[{\rm{rand}}(V), E]$, only off-site term disorder $G=[V, {\rm{rand}}(E)]$ and mixed disorder $G=[{\rm{rand}}(V), {\rm{rand}}(E)]$. Furthermore, we develop information encryption techniques within this framework.

The localization characteristics of $G=[{\rm{rand}}(V), E]$ with respect to disorder strength are analyzed in Figure \ref{fig5:anderson_cipher}a2 (discussion in Supplementary Information S5.A), at low disorder strength, the average IPR approaches zero. With the increase of disorder strength, both the maximum IPR and average IPR increase rapidly. When the disorder strength exceeds 7, the maximum IPR approaches approximately 1, indicating that the parameters are sufficient for strong localization, thereby ensuring the effective construction of a Hash map. In contrast to $G=[{\rm{rand}}(V), E]$ , $G=[V, {\rm{rand}}(E)]$ will generate multiple-to-multiple mapping, as shown in Figure \ref{fig5:anderson_cipher}b2. The identical on-site terms lead to multiple degenerate eigenenergies at zero eigenenergies with high IPR, signifying as highly localized modes. This phenomenon causes one disorder configuration mapping to multiple uncertain labels, which deviates from the Hash map. Furthermore, the localized energies extend to multiple localization positions, consequently reducing the signal-to-noise ratio (SNR) of the information. Furtherly, when both on-site and off-site terms are disordered, denoted as $G=[{\rm{rand}}(V), {\rm{rand}}(E)]$, their contributions to localization can be both synergistic and antagonistic, as Figure \ref{fig5:anderson_cipher}c2 shows. 

When the on-site disorder is weak, the localization effect induced by this term is minimal. The influence of off-site disorder is predominant; thus, increasing the off-site term will enhance localization. However, when the on-site disorder increases to a degree that can induce significant localization effects, as it is stronger than the effect of the off-site term, the off-site disorder can act as an antagonistic factor to the on-site disorder, meaning that increasing the off-site disorder will reduce or even eliminate the localization effect.

To translate the aforementioned physical properties into practical application, we have proposed and experimentally validated a physics-graph-informed Hash-based probabilistic encryption framework, as illustrated in Figure \ref{fig5:anderson_cipher}c. Unlike conventional software-based hash methods, which are primarily used for integrity verification, the hash mechanism in our framework is to map disordered sequences to specific locations. Therefore, it can also facilitate message encoding. Before encryption, the transmitter and receiver reach a consensus on a unique key (physics topology graph) and establish a protocol outlining how information will be discretized and transmitted. The information to be encrypted is allocated to the parameter space characterized by the variables of Anderson localization and is encrypted into packs (random disorder messages) using a generative AI model. For example, to encode a two-dimensional figure with the information '$\heartsuit$ TI' (symbolizing 'love topological insulator'), we build up an X-Y coordinate system labeled by the localization position and on-site disorder strength. The figure is then discretized into pixels, and each occupied pixel (in red) is encoded into an encrypted message packet.

The message packs are transmitted to the receiver for decoding, and the consensus topological system (e.g., the PTC platform) serves as the decoder. 
The data are loaded on the consensus topological system, and the disorder strength and localization position are read out. By fitting the extracted coordinates according to the established protocol, the original information is reconstructed. During this process, the transmitted messages consist solely of random numbers, and only the correct physics graph can act as the key to decode the actual information. In our experiment, the information is precisely reconstructed to its original form, with three pixels listed in Figure \ref{fig5:anderson_cipher}. For all pixels, both simulation and experiment show high consistency and fidelity (Supplementary Information S5.B). Further discussions are provided in Supplementary Information S5.C, illustrating that, in synergy with {\rm{rand}}(E) and advanced encoding formats, a dynamic physical unclonable function (PUF) \cite{gao2020physical} can be effectively implemented. In this way, the consensus key can be dynamically tuned by simultaneously setting the off-site terms, which can further increase the information safety. Beyond cryptography, such physics-informed encoding techniques can also be applied across a wide scope, including digital watermarking \cite{shih2017digital}, network security \cite{shamsoshoara2020survey}, etc. As a example, we propose a product anti‑counterfeiting scenario that leverages the integration of the Anderson physics system, the AI-fusion generative model, and electronic hardware, offering both high security and ease of use. Detailed discussion is provided in Supplementary Information S5.D.

Building on a principled message encryption framework, we have developed a practical encryption system for ASCII message delivery. At the sender (Alice), an input plaintext ASCII message of arbitrary length is first pre-processed \cite{shamir1979share,wicker1999reed} (see Method) and fed into the PGI-generative model. After transmission over a classical communication channel, the data is loaded onto the DLPTC hardware at the receiver(Bob), where the encoded strings are extracted and the decrypted message is achieved by post-processing. In experimental demonstration, we transmitted the string “HelloWorld” as well as an ASCII art depiction of the “Penrose triangle” from Southeastern University to Lanzhou University and successfully decrypted (Extended Data Fig. \ref{ext_application}). Owing to the involvement of the Anderson localization process in our approach, the encrypted messages are statistically indistinguishable from random numbers, preventing any identifiable patterns or features that could be classified or recognized, which ensures security and stealthiness. It also has strong resistance to unauthorized decryption attacks. Related analyses that confirm the high security of our system are shown in Supplementary Information S6.

\section*{Discussion and Conclusions}
The DLPTCs developed in this work provide circuit programmable flexibility, AI-boosting physics feature extraction, and closed-loop control capability. The realization of HOTIs without global symmetry and Landau levels within classical systems illustrates the platform’s programmable coupling-engineering capabilities. For the complex physics phenomena such as Anderson localization, AI-fusion research can effectively enhance the feature control and facilitate interdisciplinary exploration within the physics system. We have uncovered that multiple disorder solutions exhibit the same Anderson localization behavior, thereby revealing a compelling cryptographic application: hash-based probabilistic information encryption leveraging Anderson localization, and enable a practical message-encryption system application. The inherent reconfigurability of DLPTCs allows for various physical functionalities with a high degree of control freedom, thereby establishing them as a flexible and versatile platform without the necessity for multiple custom fabrications. The AI module empowers scientists to uncover intricate physics that are typically inaccessible through conventional methods. The integration signifies a paradigm shift with far-reaching implications, enhancing the capacity to innovate and address complex challenges across various fields. 

\section*{Methods}

\subsection*{Controllable capacitance and circuits design}
In circuit design, engineering the on-site potentials and off-site hopping terms can contribute to tuning the capacitance or inductance of the elements at the vortex and edge (connection) positions of the circuit. Normally, tunable capacitance is more practical to realize than inductance. In our circuit design, the varactor diodes are used to provide variable capacitance.

Tunable capacitors are in parallel with the vertex position contributing to on-site potential tuning, and connection lines represent the other tunable capacitors that contribute to the hopping terms. To distinguish the two class positions, we label such two types of capacitors using the symbol  $C_{V_{\bf{r}}}$ and $C_{E_{\bf{r},\bf{r'}}}$, where $\bf{r}$ represents the position of sites and $\bf{r},\bf{r'}$ represents the position of sites connected by corresponding components. Grounded inductors are in parallel with each vertex to tune the localized resonances and potentials. In this way, the circuit Laplacian is mapped to the tight-binding Hamiltonian of the topological graph. 
 
For off-site capacitance, varactor diodeis in serial with a ceramic capacitor $C_1$. The serial ceramic capacitor is used to separate the bias voltage signal to other unit cells and tune the fluctuation range of capacitance. For on-site capacitance, a Hyperabrupt Junction Tuning Varactor diode with two diodes back-to-back connected in the package is externally shunted on PCB, then in series with ceramic capacitor $C_2$ . Control voltage $V_{[\rm{control}}$ is swept by high precision programmable digital synthetic power supply from $0V$ to $10V$. 

\subsection*{Simulation and experiment}
For circuit-level simulations, open-source software LTspice is employed. 
For circuit experiments, a high-frequency impedance analyzer is used for impedance measurement and impedance spectrum scan. The in-house programmable digital synthetic power supply with 16-bit resolution is used to sweep the voltage range to construct the capacitance-bias voltage calibration curve. Python-based inverse design algorithms are integrated with the capacitance control and data acquisition instrument to constitute the fully programmable topolectrical circuit system. 
An in-house library is constituted to build the bridge between hardware driving and the interface of the GUI and the deep-learning framework, so that the hardware and software can cooperate closely.

\subsection*{Artificial intelligence modeling}
For dataset preparation, we employ the data embedding method to create physics-informed datasets that contain physics-graph information. 

For the prediction/classification of Anderson localization, the PGI-CNN classifier (AI-1 model) accepts 19×19 matrices as input. The output is a 100-dimensional softmaxed Anderson localization position probability vector. 

For the control of Anderson localization, the PGI-generator (AI-2 model) performs a U-net based diffusion network architecture and a cVAE architecture.The decoder has a structure that is the reverse of the encoder, achieved by transposed convolution layers.

To mitigate overfitting, we implement the dropout technique in the FC layer of the AI-1 model, randomly keeping 50\% of the neurons during training. The learning processes of the two models are detailed in Supplementary Information S4d, S4e, and S4g. 

All artificial intelligence models are constructed on the open-source machine learning framework PyTorch. NVIDIA GeForce RTX 4090 GPUs are utilized for model training and real-time inference.

\section*{Author contributions}
H.J., S.Y., and C.S. conceived the idea and performed the theoretical analyses. H.J. and J.H. designed the circuits and performed the experiments.  H.C. plotted the conception figures. H.J., C.S., Y.L., and T.J.C. guided the research. All the authors contributed to the discussions of the results and the preparation of the manuscript.
\section*{Data availability statement}
The datasets generated and analyzed in the current study are available from the corresponding author upon reasonable request.
\section*{Competing interests}
The authors declare no competing interests. 

\setcounter{figure}{0}
\renewcommand{\figurename}{\textbf{Extended Data Fig.}}

\begin{figure*}[p]
    \centering
    \includegraphics[width=0.9\linewidth]{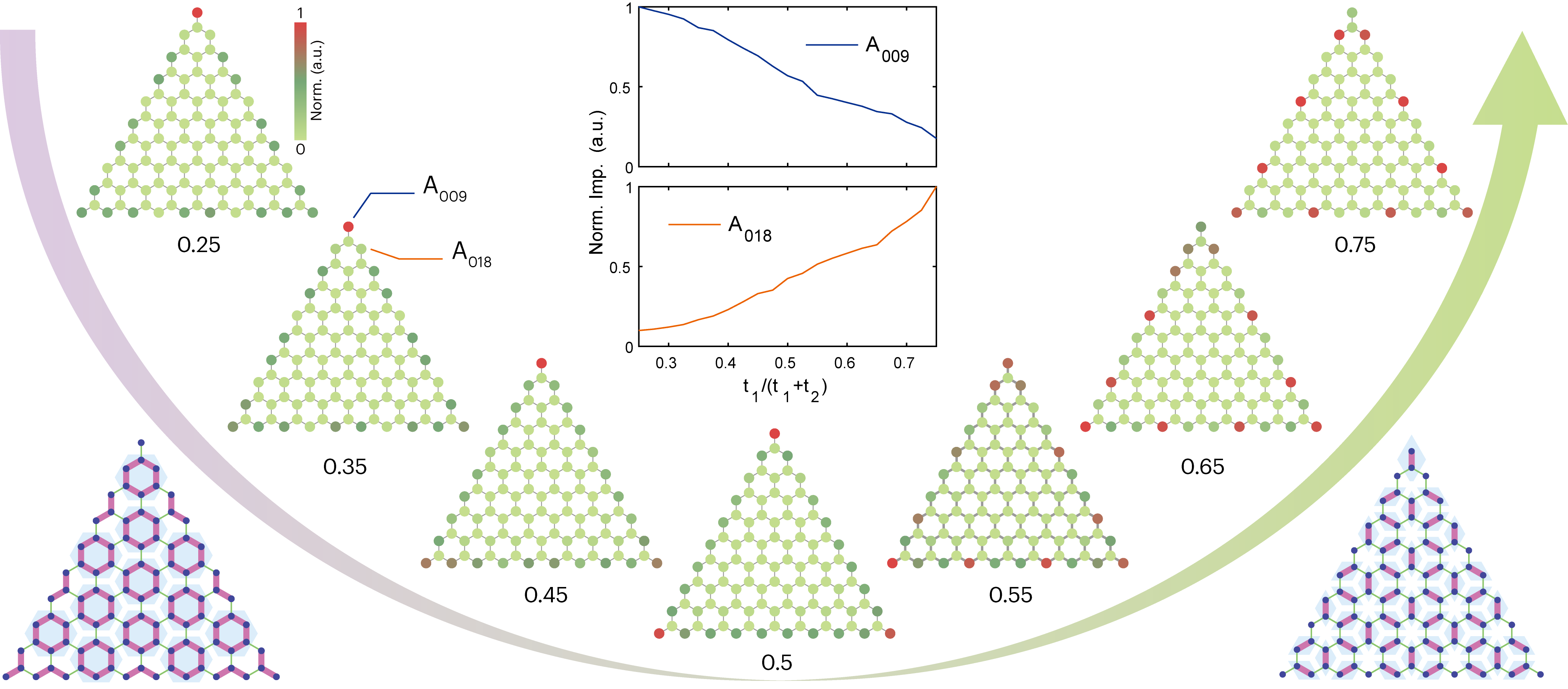}
    \caption{\textbf{Experiment observation of phase transition in HOTI without global symmetry.} By adiabatically and continuously tuning the off-site capacitance, we can experimentally observe the phase transition process. The coupling coefficient $t_1$ linearly varies from 0.25 to 0.75, while $t_2$ transitions from 0.75 to 0.25, resulting in $\gamma= t_1/(t_1+t_2)$ changing from 0.25 to 0.75. The seven-mode distribution in the figure illustrates the variation of $\gamma$ from 0.25 to 0.75. We selected two representative nodes, $A_{009}$ and $A_{018}$, to demonstrate the continuous variation of their normalized impedance, as depicted by the blue and red curves.
    }
    \label{ext_HOTI}
\end{figure*}

\begin{figure*}[p]
  \centering
  \includegraphics[width=\linewidth]{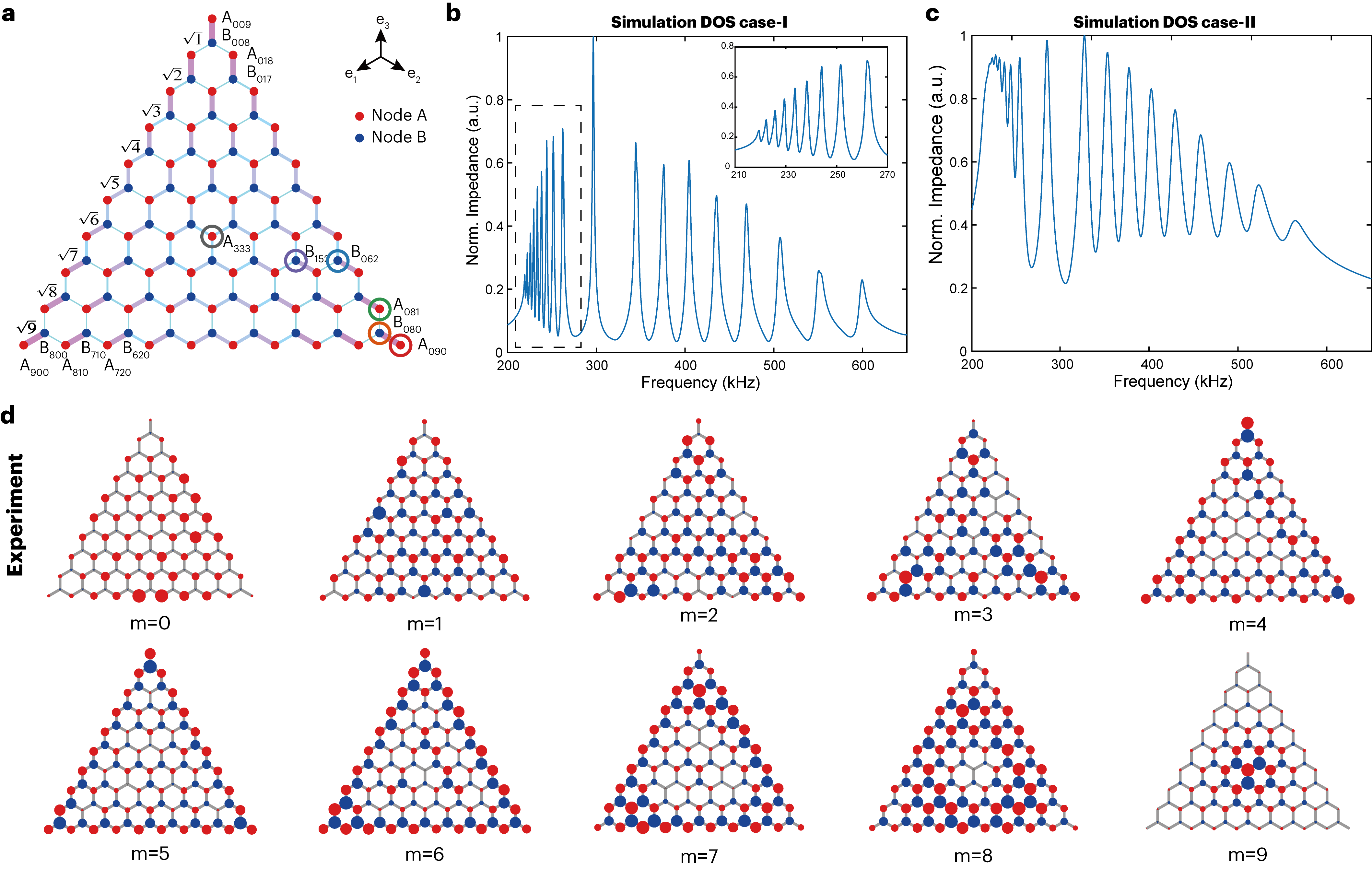}
  \caption{\textbf{Circuit implementation of Landau levels and the "breathing" mode observation.} \textbf{a}, Schematic of the flat-band lattice and representative sites at the corner, edge, and bulk position. \textbf{b}, The simulated impedance curve summation of all sites with minor parasite parameters, all 19 impedance bands can be observed, corresponding to the peaks of TBM DOS. \textbf{c}, The simulated impedance curve summation of all sites with actual parasite parameters, the lower half of the bands is compressed and closely packed. \textbf{d}, The experimental data of modes $M=1$ to $M=9$. The mode spreads from the center to the edges when mode order increases, and then shrinks to the center when continuously increasing to $M=9$. During this "breathing" process, their spatial distributions remain $C_3$ symmetry with respect to the center of the lattice.}
  \label{ext_Landau}
\end{figure*}

\newpage

\begin{figure*}[p]
  \centering
  \includegraphics[width=\linewidth]{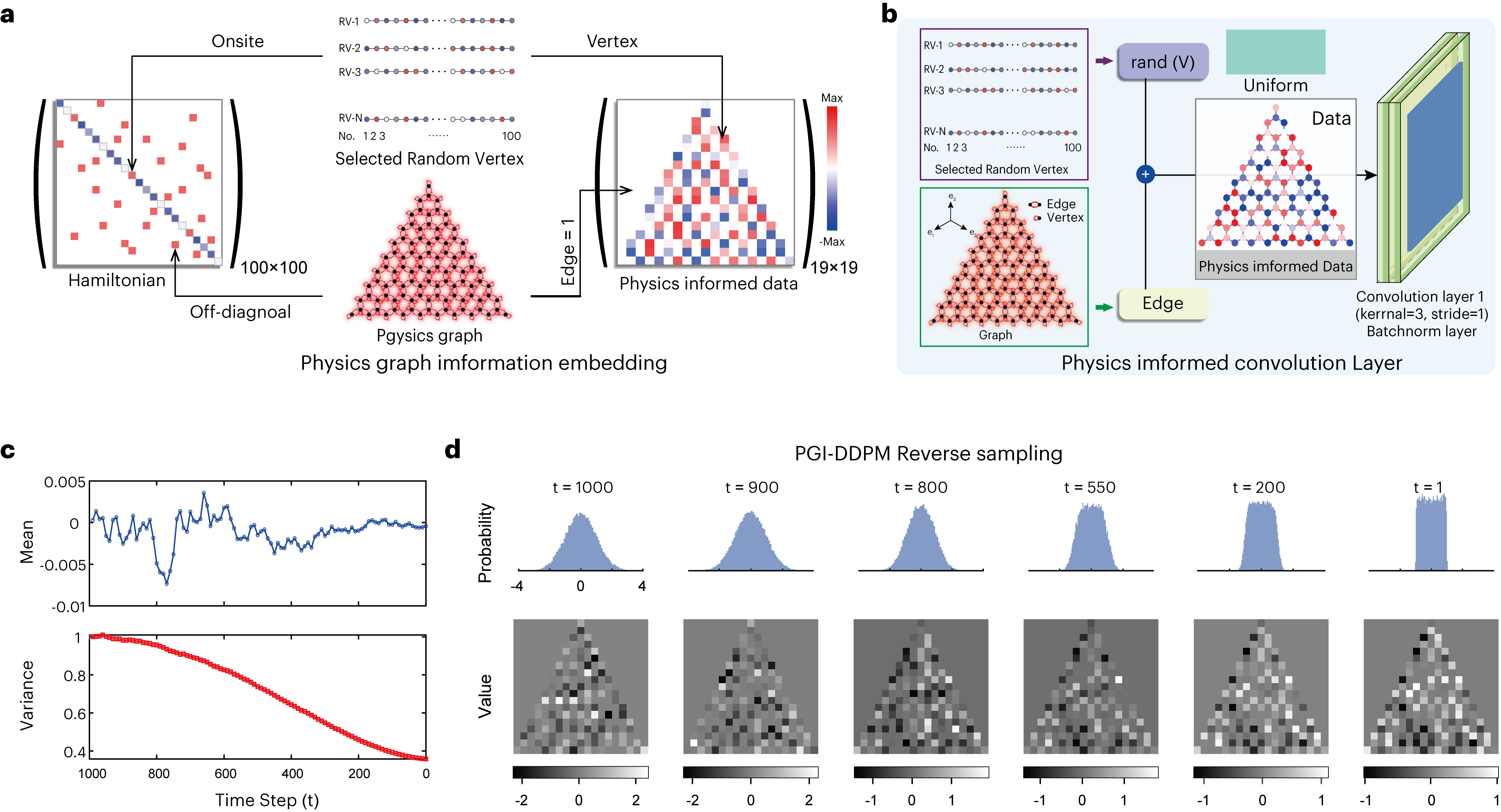}
  \caption{\textbf{Physics-graph-informed (PGI) mechanism and Controllable Anderson Localization generation process.} \textbf{a}, Schematic of physics-graph-informed dataset. \textbf{b}, Schematic of physics-graph-informed convolution layer. \textbf{c}, Evolution of mean and variance during the PGI-diffusion generation process. }
  \label{ext_AI}
\end{figure*}

\clearpage

\begin{figure*}[tp]
    \centering
    \includegraphics[width=\linewidth]{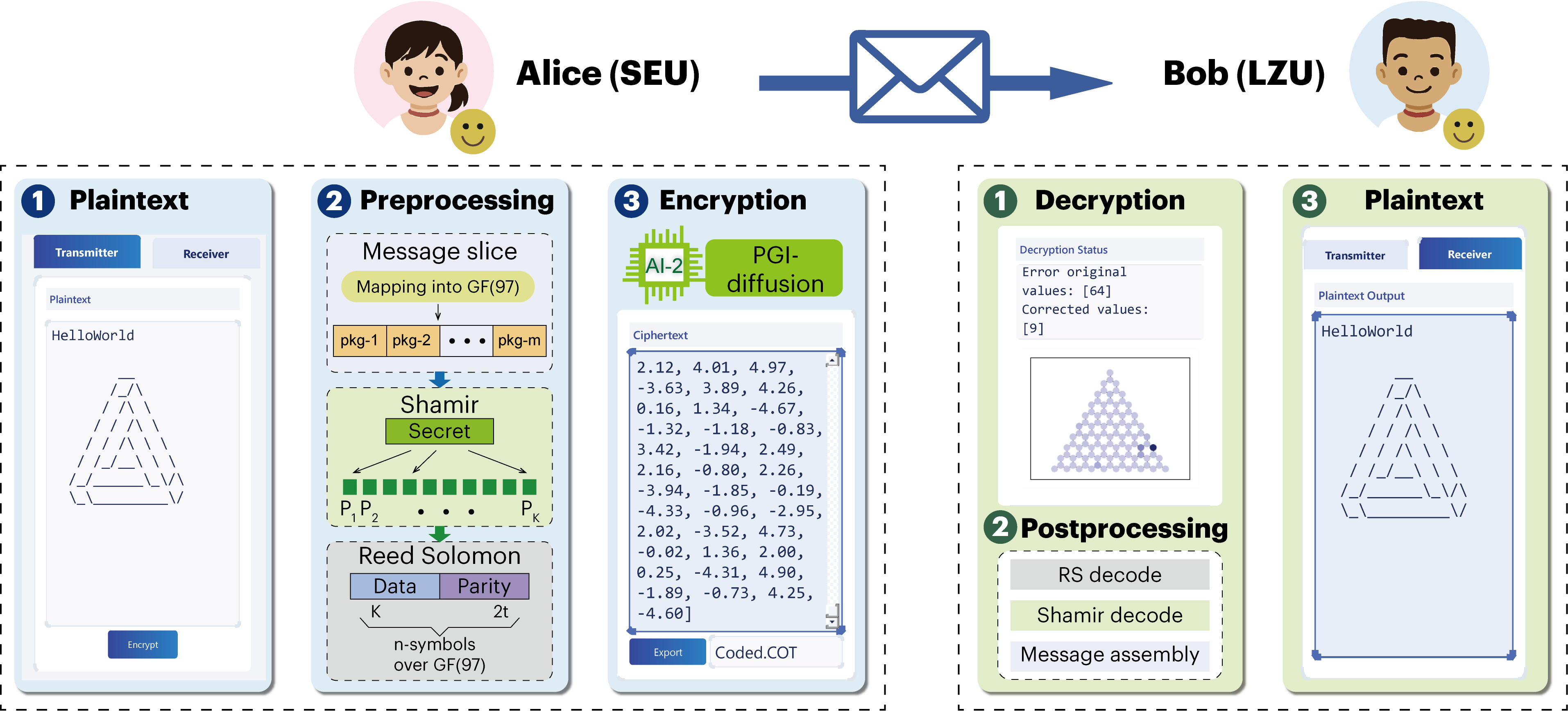}
    \caption{\textbf{The demonstration of ASCII message and ASCII art encryption and decryption.}The figure displays the ciphertext transfer steps and results for a representative ASCII message "HelloWorld" and ASCII art "Penrose Triangle". The screenshots of the key stages of our in‐house messaging system are captured and illustrated. }
    \label{ext_application}
\end{figure*}


\begin{thebibliography}{53}%
\makeatletter
\providecommand \@ifxundefined [1]{%
 \@ifx{#1\undefined}
}%
\providecommand \@ifnum [1]{%
 \ifnum #1\expandafter \@firstoftwo
 \else \expandafter \@secondoftwo
 \fi
}%
\providecommand \@ifx [1]{%
 \ifx #1\expandafter \@firstoftwo
 \else \expandafter \@secondoftwo
 \fi
}%
\providecommand \natexlab [1]{#1}%
\providecommand \enquote  [1]{``#1''}%
\providecommand \bibnamefont  [1]{#1}%
\providecommand \bibfnamefont [1]{#1}%
\providecommand \citenamefont [1]{#1}%
\providecommand \href@noop [0]{\@secondoftwo}%
\providecommand \href [0]{\begingroup \@sanitize@url \@href}%
\providecommand \@href[1]{\@@startlink{#1}\@@href}%
\providecommand \@@href[1]{\endgroup#1\@@endlink}%
\providecommand \@sanitize@url [0]{\catcode `\\12\catcode `\$12\catcode `\&12\catcode `\#12\catcode `\^12\catcode `\_12\catcode `\%12\relax}%
\providecommand \@@startlink[1]{}%
\providecommand \@@endlink[0]{}%
\providecommand \url  [0]{\begingroup\@sanitize@url \@url }%
\providecommand \@url [1]{\endgroup\@href {#1}{\urlprefix }}%
\providecommand \urlprefix  [0]{URL }%
\providecommand \Eprint [0]{\href }%
\providecommand \doibase [0]{http://dx.doi.org/}%
\providecommand \selectlanguage [0]{\@gobble}%
\providecommand \bibinfo  [0]{\@secondoftwo}%
\providecommand \bibfield  [0]{\@secondoftwo}%
\providecommand \translation [1]{[#1]}%
\providecommand \BibitemOpen [0]{}%
\providecommand \bibitemStop [0]{}%
\providecommand \bibitemNoStop [0]{.\EOS\space}%
\providecommand \EOS [0]{\spacefactor3000\relax}%
\providecommand \BibitemShut  [1]{\csname bibitem#1\endcsname}%
\let\auto@bib@innerbib\@empty
\bibitem [{\citenamefont {Feynman}(2018)}]{feynman2018simulating}%
  \BibitemOpen
  \bibfield  {author} {\bibinfo {author} {\bibfnamefont {R.~P.}\ \bibnamefont {Feynman}},\ }\href {\doibase 10.1201/9781003358817} {\emph {\bibinfo {title} {Feynman Lectures on Computation}}}\ (\bibinfo  {publisher} {CRC Press},\ \bibinfo {address} {Boca Raton},\ \bibinfo {year} {2018})\BibitemShut {NoStop}%
\bibitem [{\citenamefont {Cooper}\ \emph {et~al.}(2019)\citenamefont {Cooper}, \citenamefont {Dalibard},\ and\ \citenamefont {Spielman}}]{cooper2019topological}%
  \BibitemOpen
  \bibfield  {author} {\bibinfo {author} {\bibfnamefont {N.}~\bibnamefont {Cooper}}, \bibinfo {author} {\bibfnamefont {J.}~\bibnamefont {Dalibard}}, \ and\ \bibinfo {author} {\bibfnamefont {I.}~\bibnamefont {Spielman}},\ }\bibfield  {title} {\enquote {\bibinfo {title} {Topological bands for ultracold atoms},}\ }\href {https://doi.org/10.1103/RevModPhys.91.015005} {\bibfield  {journal} {\bibinfo  {journal} {Rev. Mod. Phys.}\ }\textbf {\bibinfo {volume} {91}},\ \bibinfo {pages} {015005} (\bibinfo {year} {2019})}\BibitemShut {NoStop}%
\bibitem [{\citenamefont {Shao}\ \emph {et~al.}(2024)\citenamefont {Shao}, \citenamefont {Wang}, \citenamefont {Zhu}, \citenamefont {Zhu}, \citenamefont {Sun}, \citenamefont {Chen}, \citenamefont {Zhang}, \citenamefont {Fan}, \citenamefont {Deng}, \citenamefont {Yao}, \citenamefont {Chen},\ and\ \citenamefont {Pan}}]{Shao2024}%
  \BibitemOpen
  \bibfield  {author} {\bibinfo {author} {\bibfnamefont {H.-J.}\ \bibnamefont {Shao}}, \bibinfo {author} {\bibfnamefont {Y.-X.}\ \bibnamefont {Wang}}, \bibinfo {author} {\bibfnamefont {D.-Z.}\ \bibnamefont {Zhu}}, \bibinfo {author} {\bibfnamefont {Y.-S.}\ \bibnamefont {Zhu}}, \bibinfo {author} {\bibfnamefont {H.-N.}\ \bibnamefont {Sun}}, \bibinfo {author} {\bibfnamefont {S.-Y.}\ \bibnamefont {Chen}}, \bibinfo {author} {\bibfnamefont {C.}~\bibnamefont {Zhang}}, \bibinfo {author} {\bibfnamefont {Z.-J.}\ \bibnamefont {Fan}}, \bibinfo {author} {\bibfnamefont {Y.}~\bibnamefont {Deng}}, \bibinfo {author} {\bibfnamefont {X.-C.}\ \bibnamefont {Yao}}, \bibinfo {author} {\bibfnamefont {Y.-A.}\ \bibnamefont {Chen}}, \ and\ \bibinfo {author} {\bibfnamefont {J.-W.}\ \bibnamefont {Pan}},\ }\bibfield  {title} {\enquote {\bibinfo {title} {Antiferromagnetic phase transition in a 3d fermionic hubbard model},}\ }\href {\doibase 10.1038/s41586-024-07689-2} {\bibfield  {journal} {\bibinfo  {journal} {Nature}\ }\textbf {\bibinfo
  {volume} {632}},\ \bibinfo {pages} {267–272} (\bibinfo {year} {2024})}\BibitemShut {NoStop}%
\bibitem [{\citenamefont {Bogaerts}\ \emph {et~al.}(2020)\citenamefont {Bogaerts}, \citenamefont {Pérez}, \citenamefont {Capmany}, \citenamefont {Miller}, \citenamefont {Poon}, \citenamefont {Englund}, \citenamefont {Morichetti},\ and\ \citenamefont {Melloni}}]{Bogaerts2020}%
  \BibitemOpen
  \bibfield  {author} {\bibinfo {author} {\bibfnamefont {W.}~\bibnamefont {Bogaerts}}, \bibinfo {author} {\bibfnamefont {D.}~\bibnamefont {Pérez}}, \bibinfo {author} {\bibfnamefont {J.}~\bibnamefont {Capmany}}, \bibinfo {author} {\bibfnamefont {D.~A.~B.}\ \bibnamefont {Miller}}, \bibinfo {author} {\bibfnamefont {J.}~\bibnamefont {Poon}}, \bibinfo {author} {\bibfnamefont {D.}~\bibnamefont {Englund}}, \bibinfo {author} {\bibfnamefont {F.}~\bibnamefont {Morichetti}}, \ and\ \bibinfo {author} {\bibfnamefont {A.}~\bibnamefont {Melloni}},\ }\bibfield  {title} {\enquote {\bibinfo {title} {Programmable photonic circuits},}\ }\href {\doibase 10.1038/s41586-020-2764-0} {\bibfield  {journal} {\bibinfo  {journal} {Nature}\ }\textbf {\bibinfo {volume} {586}},\ \bibinfo {pages} {207–216} (\bibinfo {year} {2020})}\BibitemShut {NoStop}%
\bibitem [{\citenamefont {Ozawa}\ \emph {et~al.}(2019)\citenamefont {Ozawa}, \citenamefont {Price}, \citenamefont {Amo}, \citenamefont {Goldman}, \citenamefont {Hafezi}, \citenamefont {Lu}, \citenamefont {Rechtsman}, \citenamefont {Schuster}, \citenamefont {Simon}, \citenamefont {Zilberberg},\ and\ \citenamefont {Carusotto}}]{RevModPhys.91.015006}%
  \BibitemOpen
  \bibfield  {author} {\bibinfo {author} {\bibfnamefont {T.}~\bibnamefont {Ozawa}}, \bibinfo {author} {\bibfnamefont {H.~M.}\ \bibnamefont {Price}}, \bibinfo {author} {\bibfnamefont {A.}~\bibnamefont {Amo}}, \bibinfo {author} {\bibfnamefont {N.}~\bibnamefont {Goldman}}, \bibinfo {author} {\bibfnamefont {M.}~\bibnamefont {Hafezi}}, \bibinfo {author} {\bibfnamefont {L.}~\bibnamefont {Lu}}, \bibinfo {author} {\bibfnamefont {M.~C.}\ \bibnamefont {Rechtsman}}, \bibinfo {author} {\bibfnamefont {D.}~\bibnamefont {Schuster}}, \bibinfo {author} {\bibfnamefont {J.}~\bibnamefont {Simon}}, \bibinfo {author} {\bibfnamefont {O.}~\bibnamefont {Zilberberg}}, \ and\ \bibinfo {author} {\bibfnamefont {I.}~\bibnamefont {Carusotto}},\ }\bibfield  {title} {\enquote {\bibinfo {title} {Topological photonics},}\ }\href {\doibase 10.1103/RevModPhys.91.015006} {\bibfield  {journal} {\bibinfo  {journal} {Rev. Mod. Phys.}\ }\textbf {\bibinfo {volume} {91}},\ \bibinfo {pages} {015006} (\bibinfo {year} {2019})}\BibitemShut {NoStop}%
\bibitem [{\citenamefont {On}\ \emph {et~al.}(2024)\citenamefont {On}, \citenamefont {Ashtiani}, \citenamefont {Sanchez-Jacome}, \citenamefont {Perez-Lopez}, \citenamefont {Yoo},\ and\ \citenamefont {Blanco-Redondo}}]{On2024}%
  \BibitemOpen
  \bibfield  {author} {\bibinfo {author} {\bibfnamefont {M.~B.}\ \bibnamefont {On}}, \bibinfo {author} {\bibfnamefont {F.}~\bibnamefont {Ashtiani}}, \bibinfo {author} {\bibfnamefont {D.}~\bibnamefont {Sanchez-Jacome}}, \bibinfo {author} {\bibfnamefont {D.}~\bibnamefont {Perez-Lopez}}, \bibinfo {author} {\bibfnamefont {S.~J.~B.}\ \bibnamefont {Yoo}}, \ and\ \bibinfo {author} {\bibfnamefont {A.}~\bibnamefont {Blanco-Redondo}},\ }\bibfield  {title} {\enquote {\bibinfo {title} {Programmable integrated photonics for topological hamiltonians},}\ }\href {http://dx.doi.org/10.1038/s41467-024-44939-3} {\bibfield  {journal} {\bibinfo  {journal} {Nat. Commun.}\ }\textbf {\bibinfo {volume} {15}} (\bibinfo {year} {2024})}\BibitemShut {NoStop}%
\bibitem [{\citenamefont {Dai}\ \emph {et~al.}(2024)\citenamefont {Dai}, \citenamefont {Ma}, \citenamefont {Mao}, \citenamefont {Ao}, \citenamefont {Jia}, \citenamefont {Zheng}, \citenamefont {Zhai}, \citenamefont {Yang}, \citenamefont {Li}, \citenamefont {Tang}, \citenamefont {Luo}, \citenamefont {Zhang}, \citenamefont {Hu}, \citenamefont {Gong},\ and\ \citenamefont {Wang}}]{dai2024programmable}%
  \BibitemOpen
  \bibfield  {author} {\bibinfo {author} {\bibfnamefont {T.}~\bibnamefont {Dai}}, \bibinfo {author} {\bibfnamefont {A.}~\bibnamefont {Ma}}, \bibinfo {author} {\bibfnamefont {J.}~\bibnamefont {Mao}}, \bibinfo {author} {\bibfnamefont {Y.}~\bibnamefont {Ao}}, \bibinfo {author} {\bibfnamefont {X.}~\bibnamefont {Jia}}, \bibinfo {author} {\bibfnamefont {Y.}~\bibnamefont {Zheng}}, \bibinfo {author} {\bibfnamefont {C.}~\bibnamefont {Zhai}}, \bibinfo {author} {\bibfnamefont {Y.}~\bibnamefont {Yang}}, \bibinfo {author} {\bibfnamefont {Z.}~\bibnamefont {Li}}, \bibinfo {author} {\bibfnamefont {B.}~\bibnamefont {Tang}}, \bibinfo {author} {\bibfnamefont {J.}~\bibnamefont {Luo}}, \bibinfo {author} {\bibfnamefont {B.}~\bibnamefont {Zhang}}, \bibinfo {author} {\bibfnamefont {X.}~\bibnamefont {Hu}}, \bibinfo {author} {\bibfnamefont {Q.}~\bibnamefont {Gong}}, \ and\ \bibinfo {author} {\bibfnamefont {J.}~\bibnamefont {Wang}},\ }\bibfield  {title} {\enquote {\bibinfo {title} {A programmable topological photonic chip},}\ }\href
  {http://dx.doi.org/10.1038/s41563-024-01904-1} {\bibfield  {journal} {\bibinfo  {journal} {Nat. Mater.}\ }\textbf {\bibinfo {volume} {23}} (\bibinfo {year} {2024})}\BibitemShut {NoStop}%
\bibitem [{\citenamefont {Koh}\ \emph {et~al.}(2022)\citenamefont {Koh}, \citenamefont {Tai},\ and\ \citenamefont {Lee}}]{PhysRevLett.129.140502}%
  \BibitemOpen
  \bibfield  {author} {\bibinfo {author} {\bibfnamefont {J.~M.}\ \bibnamefont {Koh}}, \bibinfo {author} {\bibfnamefont {T.}~\bibnamefont {Tai}}, \ and\ \bibinfo {author} {\bibfnamefont {C.~H.}\ \bibnamefont {Lee}},\ }\bibfield  {title} {\enquote {\bibinfo {title} {Simulation of interaction-induced chiral topological dynamics on a digital quantum computer},}\ }\href {\doibase 10.1103/PhysRevLett.129.140502} {\bibfield  {journal} {\bibinfo  {journal} {Phys. Rev. Lett.}\ }\textbf {\bibinfo {volume} {129}},\ \bibinfo {pages} {140502} (\bibinfo {year} {2022})}\BibitemShut {NoStop}%
\bibitem [{\citenamefont {Koh}\ \emph {et~al.}(2024)\citenamefont {Koh}, \citenamefont {Tai},\ and\ \citenamefont {Lee}}]{koh2024realization}%
  \BibitemOpen
  \bibfield  {author} {\bibinfo {author} {\bibfnamefont {J.~M.}\ \bibnamefont {Koh}}, \bibinfo {author} {\bibfnamefont {T.}~\bibnamefont {Tai}}, \ and\ \bibinfo {author} {\bibfnamefont {C.~H.}\ \bibnamefont {Lee}},\ }\bibfield  {title} {\enquote {\bibinfo {title} {Realization of higher-order topological lattices on a quantum computer},}\ }\href {https://doi.org/10.1038/s41467-024-49648-5} {\bibfield  {journal} {\bibinfo  {journal} {Nat. Commun.}\ }\textbf {\bibinfo {volume} {15}},\ \bibinfo {pages} {5807} (\bibinfo {year} {2024})}\BibitemShut {NoStop}%
\bibitem [{\citenamefont {You}\ \emph {et~al.}(2021)\citenamefont {You}, \citenamefont {Ma}, \citenamefont {Lan}, \citenamefont {Xiao}, \citenamefont {Panoiu},\ and\ \citenamefont {Cui}}]{You2021}%
  \BibitemOpen
  \bibfield  {author} {\bibinfo {author} {\bibfnamefont {J.}~\bibnamefont {You}}, \bibinfo {author} {\bibfnamefont {Q.}~\bibnamefont {Ma}}, \bibinfo {author} {\bibfnamefont {Z.}~\bibnamefont {Lan}}, \bibinfo {author} {\bibfnamefont {Q.}~\bibnamefont {Xiao}}, \bibinfo {author} {\bibfnamefont {N.~C.}\ \bibnamefont {Panoiu}}, \ and\ \bibinfo {author} {\bibfnamefont {T.}~\bibnamefont {Cui}},\ }\bibfield  {title} {\enquote {\bibinfo {title} {Reprogrammable plasmonic topological insulators with ultrafast control},}\ }\href {http://dx.doi.org/10.1038/s41467-021-25835-6} {\bibfield  {journal} {\bibinfo  {journal} {Nat. Commun.}\ }\textbf {\bibinfo {volume} {12}} (\bibinfo {year} {2021})}\BibitemShut {NoStop}%
\bibitem [{\citenamefont {Wu}\ \emph {et~al.}(2021)\citenamefont {Wu}, \citenamefont {Bao}, \citenamefont {Cao}, \citenamefont {Chen}, \citenamefont {Chen}, \citenamefont {Chen}, \citenamefont {Chung}, \citenamefont {Deng}, \citenamefont {Du}, \citenamefont {Fan}, \citenamefont {Gong}, \citenamefont {Guo}, \citenamefont {Guo}, \citenamefont {Guo}, \citenamefont {Han}, \citenamefont {Hong}, \citenamefont {Huang}, \citenamefont {Huo}, \citenamefont {Li}, \citenamefont {Li}, \citenamefont {Li}, \citenamefont {Li}, \citenamefont {Liang}, \citenamefont {Lin}, \citenamefont {Lin}, \citenamefont {Qian}, \citenamefont {Qiao}, \citenamefont {Rong}, \citenamefont {Su}, \citenamefont {Sun}, \citenamefont {Wang}, \citenamefont {Wang}, \citenamefont {Wu}, \citenamefont {Xu}, \citenamefont {Yan}, \citenamefont {Yang}, \citenamefont {Yang}, \citenamefont {Ye}, \citenamefont {Yin}, \citenamefont {Ying}, \citenamefont {Yu}, \citenamefont {Zha}, \citenamefont {Zhang}, \citenamefont {Zhang}, \citenamefont {Zhang}, \citenamefont
  {Zhang}, \citenamefont {Zhao}, \citenamefont {Zhao}, \citenamefont {Zhou}, \citenamefont {Zhu}, \citenamefont {Lu}, \citenamefont {Peng}, \citenamefont {Zhu},\ and\ \citenamefont {Pan}}]{PhysRevLett.127.180501}%
  \BibitemOpen
  \bibfield  {author} {\bibinfo {author} {\bibfnamefont {Y.}~\bibnamefont {Wu}}, \bibinfo {author} {\bibfnamefont {W.-S.}\ \bibnamefont {Bao}}, \bibinfo {author} {\bibfnamefont {S.}~\bibnamefont {Cao}}, \bibinfo {author} {\bibfnamefont {F.}~\bibnamefont {Chen}}, \bibinfo {author} {\bibfnamefont {M.-C.}\ \bibnamefont {Chen}}, \bibinfo {author} {\bibfnamefont {X.}~\bibnamefont {Chen}}, \bibinfo {author} {\bibfnamefont {T.-H.}\ \bibnamefont {Chung}}, \bibinfo {author} {\bibfnamefont {H.}~\bibnamefont {Deng}}, \bibinfo {author} {\bibfnamefont {Y.}~\bibnamefont {Du}}, \bibinfo {author} {\bibfnamefont {D.}~\bibnamefont {Fan}}, \bibinfo {author} {\bibfnamefont {M.}~\bibnamefont {Gong}}, \bibinfo {author} {\bibfnamefont {C.}~\bibnamefont {Guo}}, \bibinfo {author} {\bibfnamefont {C.}~\bibnamefont {Guo}}, \bibinfo {author} {\bibfnamefont {S.}~\bibnamefont {Guo}}, \bibinfo {author} {\bibfnamefont {L.}~\bibnamefont {Han}}, \bibinfo {author} {\bibfnamefont {L.}~\bibnamefont {Hong}}, \bibinfo {author} {\bibfnamefont
  {H.-L.}\ \bibnamefont {Huang}}, \bibinfo {author} {\bibfnamefont {Y.-H.}\ \bibnamefont {Huo}}, \bibinfo {author} {\bibfnamefont {L.}~\bibnamefont {Li}}, \bibinfo {author} {\bibfnamefont {N.}~\bibnamefont {Li}}, \bibinfo {author} {\bibfnamefont {S.}~\bibnamefont {Li}}, \bibinfo {author} {\bibfnamefont {Y.}~\bibnamefont {Li}}, \bibinfo {author} {\bibfnamefont {F.}~\bibnamefont {Liang}}, \bibinfo {author} {\bibfnamefont {C.}~\bibnamefont {Lin}}, \bibinfo {author} {\bibfnamefont {J.}~\bibnamefont {Lin}}, \bibinfo {author} {\bibfnamefont {H.}~\bibnamefont {Qian}}, \bibinfo {author} {\bibfnamefont {D.}~\bibnamefont {Qiao}}, \bibinfo {author} {\bibfnamefont {H.}~\bibnamefont {Rong}}, \bibinfo {author} {\bibfnamefont {H.}~\bibnamefont {Su}}, \bibinfo {author} {\bibfnamefont {L.}~\bibnamefont {Sun}}, \bibinfo {author} {\bibfnamefont {L.}~\bibnamefont {Wang}}, \bibinfo {author} {\bibfnamefont {S.}~\bibnamefont {Wang}}, \bibinfo {author} {\bibfnamefont {D.}~\bibnamefont {Wu}}, \bibinfo {author} {\bibfnamefont
  {Y.}~\bibnamefont {Xu}}, \bibinfo {author} {\bibfnamefont {K.}~\bibnamefont {Yan}}, \bibinfo {author} {\bibfnamefont {W.}~\bibnamefont {Yang}}, \bibinfo {author} {\bibfnamefont {Y.}~\bibnamefont {Yang}}, \bibinfo {author} {\bibfnamefont {Y.}~\bibnamefont {Ye}}, \bibinfo {author} {\bibfnamefont {J.}~\bibnamefont {Yin}}, \bibinfo {author} {\bibfnamefont {C.}~\bibnamefont {Ying}}, \bibinfo {author} {\bibfnamefont {J.}~\bibnamefont {Yu}}, \bibinfo {author} {\bibfnamefont {C.}~\bibnamefont {Zha}}, \bibinfo {author} {\bibfnamefont {C.}~\bibnamefont {Zhang}}, \bibinfo {author} {\bibfnamefont {H.}~\bibnamefont {Zhang}}, \bibinfo {author} {\bibfnamefont {K.}~\bibnamefont {Zhang}}, \bibinfo {author} {\bibfnamefont {Y.}~\bibnamefont {Zhang}}, \bibinfo {author} {\bibfnamefont {H.}~\bibnamefont {Zhao}}, \bibinfo {author} {\bibfnamefont {Y.}~\bibnamefont {Zhao}}, \bibinfo {author} {\bibfnamefont {L.}~\bibnamefont {Zhou}}, \bibinfo {author} {\bibfnamefont {Q.}~\bibnamefont {Zhu}}, \bibinfo {author} {\bibfnamefont {C.-Y.}\
  \bibnamefont {Lu}}, \bibinfo {author} {\bibfnamefont {C.-Z.}\ \bibnamefont {Peng}}, \bibinfo {author} {\bibfnamefont {X.}~\bibnamefont {Zhu}}, \ and\ \bibinfo {author} {\bibfnamefont {J.-W.}\ \bibnamefont {Pan}},\ }\bibfield  {title} {\enquote {\bibinfo {title} {Strong quantum computational advantage using a superconducting quantum processor},}\ }\href {\doibase 10.1103/PhysRevLett.127.180501} {\bibfield  {journal} {\bibinfo  {journal} {Phys. Rev. Lett.}\ }\textbf {\bibinfo {volume} {127}},\ \bibinfo {pages} {180501} (\bibinfo {year} {2021})}\BibitemShut {NoStop}%
\bibitem [{\citenamefont {Xiao}\ \emph {et~al.}(2010)\citenamefont {Xiao}, \citenamefont {Chang},\ and\ \citenamefont {Niu}}]{RevModPhys.82.1959}%
  \BibitemOpen
  \bibfield  {author} {\bibinfo {author} {\bibfnamefont {D.}~\bibnamefont {Xiao}}, \bibinfo {author} {\bibfnamefont {M.-C.}\ \bibnamefont {Chang}}, \ and\ \bibinfo {author} {\bibfnamefont {Q.}~\bibnamefont {Niu}},\ }\bibfield  {title} {\enquote {\bibinfo {title} {Berry phase effects on electronic properties},}\ }\href {\doibase 10.1103/RevModPhys.82.1959} {\bibfield  {journal} {\bibinfo  {journal} {Rev. Mod. Phys.}\ }\textbf {\bibinfo {volume} {82}},\ \bibinfo {pages} {1959} (\bibinfo {year} {2010})}\BibitemShut {NoStop}%
\bibitem [{\citenamefont {Qi}\ and\ \citenamefont {Zhang}(2011)}]{RevModPhys.83.1057}%
  \BibitemOpen
  \bibfield  {author} {\bibinfo {author} {\bibfnamefont {X.-L.}\ \bibnamefont {Qi}}\ and\ \bibinfo {author} {\bibfnamefont {S.-C.}\ \bibnamefont {Zhang}},\ }\bibfield  {title} {\enquote {\bibinfo {title} {Topological insulators and superconductors},}\ }\href {\doibase 10.1103/RevModPhys.83.1057} {\bibfield  {journal} {\bibinfo  {journal} {Rev. Mod. Phys.}\ }\textbf {\bibinfo {volume} {83}},\ \bibinfo {pages} {1057} (\bibinfo {year} {2011})}\BibitemShut {NoStop}%
\bibitem [{\citenamefont {Benalcazar}\ \emph {et~al.}(2017)\citenamefont {Benalcazar}, \citenamefont {Bernevig},\ and\ \citenamefont {Hughes}}]{Benalcazar61}%
  \BibitemOpen
  \bibfield  {author} {\bibinfo {author} {\bibfnamefont {W.~A.}\ \bibnamefont {Benalcazar}}, \bibinfo {author} {\bibfnamefont {B.~A.}\ \bibnamefont {Bernevig}}, \ and\ \bibinfo {author} {\bibfnamefont {T.~L.}\ \bibnamefont {Hughes}},\ }\bibfield  {title} {\enquote {\bibinfo {title} {Quantized electric multipole insulators},}\ }\href {\doibase 10.1126/science.aah6442} {\bibfield  {journal} {\bibinfo  {journal} {Science}\ }\textbf {\bibinfo {volume} {357}},\ \bibinfo {pages} {61} (\bibinfo {year} {2017})}\BibitemShut {NoStop}%
\bibitem [{\citenamefont {Rechtsman}\ \emph {et~al.}(2013)\citenamefont {Rechtsman}, \citenamefont {Zeuner}, \citenamefont {T{\"u}nnermann}, \citenamefont {Nolte}, \citenamefont {Segev},\ and\ \citenamefont {Szameit}}]{rechtsman2013strain}%
  \BibitemOpen
  \bibfield  {author} {\bibinfo {author} {\bibfnamefont {M.~C.}\ \bibnamefont {Rechtsman}}, \bibinfo {author} {\bibfnamefont {J.~M.}\ \bibnamefont {Zeuner}}, \bibinfo {author} {\bibfnamefont {A.}~\bibnamefont {T{\"u}nnermann}}, \bibinfo {author} {\bibfnamefont {S.}~\bibnamefont {Nolte}}, \bibinfo {author} {\bibfnamefont {M.}~\bibnamefont {Segev}}, \ and\ \bibinfo {author} {\bibfnamefont {A.}~\bibnamefont {Szameit}},\ }\bibfield  {title} {\enquote {\bibinfo {title} {Strain-induced pseudomagnetic field and photonic landau levels in dielectric structures},}\ }\href {https://doi.org/10.1038/nphoton.2012.302} {\bibfield  {journal} {\bibinfo  {journal} {Nat. Photonics}\ }\textbf {\bibinfo {volume} {7}},\ \bibinfo {pages} {153} (\bibinfo {year} {2013})}\BibitemShut {NoStop}%
\bibitem [{\citenamefont {Yang}\ \emph {et~al.}(2017)\citenamefont {Yang}, \citenamefont {Gao}, \citenamefont {Yang},\ and\ \citenamefont {Zhang}}]{PhysRevLett.118.194301}%
  \BibitemOpen
  \bibfield  {author} {\bibinfo {author} {\bibfnamefont {Z.}~\bibnamefont {Yang}}, \bibinfo {author} {\bibfnamefont {F.}~\bibnamefont {Gao}}, \bibinfo {author} {\bibfnamefont {Y.}~\bibnamefont {Yang}}, \ and\ \bibinfo {author} {\bibfnamefont {B.}~\bibnamefont {Zhang}},\ }\bibfield  {title} {\enquote {\bibinfo {title} {Strain-induced gauge field and landau levels in acoustic structures},}\ }\href {\doibase 10.1103/PhysRevLett.118.194301} {\bibfield  {journal} {\bibinfo  {journal} {Phys. Rev. Lett.}\ }\textbf {\bibinfo {volume} {118}},\ \bibinfo {pages} {194301} (\bibinfo {year} {2017})}\BibitemShut {NoStop}%
\bibitem [{\citenamefont {Barsukova}\ \emph {et~al.}(2024)\citenamefont {Barsukova}, \citenamefont {Gris{\'e}}, \citenamefont {Zhang}, \citenamefont {Vaidya}, \citenamefont {Guglielmon}, \citenamefont {Weinstein}, \citenamefont {He}, \citenamefont {Zhen}, \citenamefont {McEntaffer},\ and\ \citenamefont {Rechtsman}}]{barsukova2024direct}%
  \BibitemOpen
  \bibfield  {author} {\bibinfo {author} {\bibfnamefont {M.}~\bibnamefont {Barsukova}}, \bibinfo {author} {\bibfnamefont {F.}~\bibnamefont {Gris{\'e}}}, \bibinfo {author} {\bibfnamefont {Z.}~\bibnamefont {Zhang}}, \bibinfo {author} {\bibfnamefont {S.}~\bibnamefont {Vaidya}}, \bibinfo {author} {\bibfnamefont {J.}~\bibnamefont {Guglielmon}}, \bibinfo {author} {\bibfnamefont {M.~I.}\ \bibnamefont {Weinstein}}, \bibinfo {author} {\bibfnamefont {L.}~\bibnamefont {He}}, \bibinfo {author} {\bibfnamefont {B.}~\bibnamefont {Zhen}}, \bibinfo {author} {\bibfnamefont {R.}~\bibnamefont {McEntaffer}}, \ and\ \bibinfo {author} {\bibfnamefont {M.~C.}\ \bibnamefont {Rechtsman}},\ }\bibfield  {title} {\enquote {\bibinfo {title} {Direct observation of landau levels in silicon photonic crystals},}\ }\href {https://doi.org/10.1038/s41566-024-01425-y} {\bibfield  {journal} {\bibinfo  {journal} {Nat. Photonics}\ }\textbf {\bibinfo {volume} {18}},\ \bibinfo {pages} {580} (\bibinfo {year} {2024})}\BibitemShut {NoStop}%
\bibitem [{\citenamefont {Segev}\ \emph {et~al.}(2013)\citenamefont {Segev}, \citenamefont {Silberberg},\ and\ \citenamefont {Christodoulides}}]{Segev2013}%
  \BibitemOpen
  \bibfield  {author} {\bibinfo {author} {\bibfnamefont {M.}~\bibnamefont {Segev}}, \bibinfo {author} {\bibfnamefont {Y.}~\bibnamefont {Silberberg}}, \ and\ \bibinfo {author} {\bibfnamefont {D.~N.}\ \bibnamefont {Christodoulides}},\ }\bibfield  {title} {\enquote {\bibinfo {title} {Anderson localization of light},}\ }\href {\doibase 10.1038/nphoton.2013.30} {\bibfield  {journal} {\bibinfo  {journal} {Nat. Photonics}\ }\textbf {\bibinfo {volume} {7}},\ \bibinfo {pages} {197–204} (\bibinfo {year} {2013})}\BibitemShut {NoStop}%
\bibitem [{\citenamefont {Yang}\ \emph {et~al.}(2024{\natexlab{a}})\citenamefont {Yang}, \citenamefont {Chapman}, \citenamefont {Haylock}, \citenamefont {Lenzini}, \citenamefont {Joglekar}, \citenamefont {Lobino},\ and\ \citenamefont {Peruzzo}}]{Yang2024}%
  \BibitemOpen
  \bibfield  {author} {\bibinfo {author} {\bibfnamefont {Y.}~\bibnamefont {Yang}}, \bibinfo {author} {\bibfnamefont {R.~J.}\ \bibnamefont {Chapman}}, \bibinfo {author} {\bibfnamefont {B.}~\bibnamefont {Haylock}}, \bibinfo {author} {\bibfnamefont {F.}~\bibnamefont {Lenzini}}, \bibinfo {author} {\bibfnamefont {Y.~N.}\ \bibnamefont {Joglekar}}, \bibinfo {author} {\bibfnamefont {M.}~\bibnamefont {Lobino}}, \ and\ \bibinfo {author} {\bibfnamefont {A.}~\bibnamefont {Peruzzo}},\ }\bibfield  {title} {\enquote {\bibinfo {title} {Programmable high-dimensional hamiltonian in a photonic waveguide array},}\ }\href {\doibase 10.1038/s41467-023-44185-z} {\bibfield  {journal} {\bibinfo  {journal} {Nat. Commun.}\ }\textbf {\bibinfo {volume} {15}} (\bibinfo {year} {2024}{\natexlab{a}}),\ 10.1038/s41467-023-44185-z}\BibitemShut {NoStop}%
\bibitem [{\citenamefont {Lee}\ \emph {et~al.}(2018{\natexlab{a}})\citenamefont {Lee}, \citenamefont {Imhof}, \citenamefont {Berger}, \citenamefont {Bayer}, \citenamefont {Brehm}, \citenamefont {Molenkamp}, \citenamefont {Kiessling},\ and\ \citenamefont {Thomale}}]{Lee2018}%
  \BibitemOpen
  \bibfield  {author} {\bibinfo {author} {\bibfnamefont {C.~H.}\ \bibnamefont {Lee}}, \bibinfo {author} {\bibfnamefont {S.}~\bibnamefont {Imhof}}, \bibinfo {author} {\bibfnamefont {C.}~\bibnamefont {Berger}}, \bibinfo {author} {\bibfnamefont {F.}~\bibnamefont {Bayer}}, \bibinfo {author} {\bibfnamefont {J.}~\bibnamefont {Brehm}}, \bibinfo {author} {\bibfnamefont {L.~W.}\ \bibnamefont {Molenkamp}}, \bibinfo {author} {\bibfnamefont {T.}~\bibnamefont {Kiessling}}, \ and\ \bibinfo {author} {\bibfnamefont {R.}~\bibnamefont {Thomale}},\ }\bibfield  {title} {\enquote {\bibinfo {title} {Topolectrical circuits},}\ }\href {https://doi.org/10.1038/s42005-018-0035-2} {\bibfield  {journal} {\bibinfo  {journal} {Commun. Phys.}\ }\textbf {\bibinfo {volume} {1}},\ \bibinfo {pages} {39} (\bibinfo {year} {2018}{\natexlab{a}})}\BibitemShut {NoStop}%
\bibitem [{\citenamefont {Hadad}\ \emph {et~al.}(2018)\citenamefont {Hadad}, \citenamefont {Soric}, \citenamefont {Khanikaev},\ and\ \citenamefont {Alù}}]{Hadad2018}%
  \BibitemOpen
  \bibfield  {author} {\bibinfo {author} {\bibfnamefont {Y.}~\bibnamefont {Hadad}}, \bibinfo {author} {\bibfnamefont {J.~C.}\ \bibnamefont {Soric}}, \bibinfo {author} {\bibfnamefont {A.~B.}\ \bibnamefont {Khanikaev}}, \ and\ \bibinfo {author} {\bibfnamefont {A.}~\bibnamefont {Alù}},\ }\bibfield  {title} {\enquote {\bibinfo {title} {Self-induced topological protection in nonlinear circuit arrays},}\ }\href {\doibase 10.1038/s41928-018-0042-z} {\bibfield  {journal} {\bibinfo  {journal} {Nat. Electron.}\ }\textbf {\bibinfo {volume} {1}},\ \bibinfo {pages} {178} (\bibinfo {year} {2018})}\BibitemShut {NoStop}%
\bibitem [{\citenamefont {Wu}\ \emph {et~al.}(2022)\citenamefont {Wu}, \citenamefont {Wang}, \citenamefont {Biao}, \citenamefont {Fei}, \citenamefont {Zhang}, \citenamefont {Yin}, \citenamefont {Hu}, \citenamefont {Song}, \citenamefont {Wu}, \citenamefont {Song},\ and\ \citenamefont {Yu}}]{Wu2022}%
  \BibitemOpen
  \bibfield  {author} {\bibinfo {author} {\bibfnamefont {J.}~\bibnamefont {Wu}}, \bibinfo {author} {\bibfnamefont {Z.}~\bibnamefont {Wang}}, \bibinfo {author} {\bibfnamefont {Y.}~\bibnamefont {Biao}}, \bibinfo {author} {\bibfnamefont {F.}~\bibnamefont {Fei}}, \bibinfo {author} {\bibfnamefont {S.}~\bibnamefont {Zhang}}, \bibinfo {author} {\bibfnamefont {Z.}~\bibnamefont {Yin}}, \bibinfo {author} {\bibfnamefont {Y.}~\bibnamefont {Hu}}, \bibinfo {author} {\bibfnamefont {Z.}~\bibnamefont {Song}}, \bibinfo {author} {\bibfnamefont {T.}~\bibnamefont {Wu}}, \bibinfo {author} {\bibfnamefont {F.}~\bibnamefont {Song}}, \ and\ \bibinfo {author} {\bibfnamefont {R.}~\bibnamefont {Yu}},\ }\bibfield  {title} {\enquote {\bibinfo {title} {Non-abelian gauge fields in circuit systems},}\ }\href {\doibase 10.1038/s41928-022-00833-8} {\bibfield  {journal} {\bibinfo  {journal} {Nat. Electron.}\ }\textbf {\bibinfo {volume} {5}},\ \bibinfo {pages} {635} (\bibinfo {year} {2022})}\BibitemShut {NoStop}%
\bibitem [{\citenamefont {Carleo}\ \emph {et~al.}(2019)\citenamefont {Carleo}, \citenamefont {Cirac}, \citenamefont {Cranmer}, \citenamefont {Daudet}, \citenamefont {Schuld}, \citenamefont {Tishby}, \citenamefont {Vogt-Maranto},\ and\ \citenamefont {Zdeborov{\'a}}}]{carleo2019machine}%
  \BibitemOpen
  \bibfield  {author} {\bibinfo {author} {\bibfnamefont {G.}~\bibnamefont {Carleo}}, \bibinfo {author} {\bibfnamefont {I.}~\bibnamefont {Cirac}}, \bibinfo {author} {\bibfnamefont {K.}~\bibnamefont {Cranmer}}, \bibinfo {author} {\bibfnamefont {L.}~\bibnamefont {Daudet}}, \bibinfo {author} {\bibfnamefont {M.}~\bibnamefont {Schuld}}, \bibinfo {author} {\bibfnamefont {N.}~\bibnamefont {Tishby}}, \bibinfo {author} {\bibfnamefont {L.}~\bibnamefont {Vogt-Maranto}}, \ and\ \bibinfo {author} {\bibfnamefont {L.}~\bibnamefont {Zdeborov{\'a}}},\ }\bibfield  {title} {\enquote {\bibinfo {title} {Machine learning and the physical sciences},}\ }\href {https://doi.org/10.1103/RevModPhys.91.045002} {\bibfield  {journal} {\bibinfo  {journal} {Rev. Mod. Phys.}\ }\textbf {\bibinfo {volume} {91}},\ \bibinfo {pages} {045002} (\bibinfo {year} {2019})}\BibitemShut {NoStop}%
\bibitem [{\citenamefont {Raissi}\ \emph {et~al.}(2020)\citenamefont {Raissi}, \citenamefont {Yazdani},\ and\ \citenamefont {Karniadakis}}]{raissi2020hidden}%
  \BibitemOpen
  \bibfield  {author} {\bibinfo {author} {\bibfnamefont {M.}~\bibnamefont {Raissi}}, \bibinfo {author} {\bibfnamefont {A.}~\bibnamefont {Yazdani}}, \ and\ \bibinfo {author} {\bibfnamefont {G.~E.}\ \bibnamefont {Karniadakis}},\ }\bibfield  {title} {\enquote {\bibinfo {title} {Hidden fluid mechanics: Learning velocity and pressure fields from flow visualizations},}\ }\href {https://doi.org/10.1126/science.aaw4741} {\bibfield  {journal} {\bibinfo  {journal} {Science}\ }\textbf {\bibinfo {volume} {367}},\ \bibinfo {pages} {1026} (\bibinfo {year} {2020})}\BibitemShut {NoStop}%
\bibitem [{\citenamefont {Karniadakis}\ \emph {et~al.}(2021)\citenamefont {Karniadakis}, \citenamefont {Kevrekidis}, \citenamefont {Lu}, \citenamefont {Perdikaris}, \citenamefont {Wang},\ and\ \citenamefont {Yang}}]{karniadakis2021physics}%
  \BibitemOpen
  \bibfield  {author} {\bibinfo {author} {\bibfnamefont {G.~E.}\ \bibnamefont {Karniadakis}}, \bibinfo {author} {\bibfnamefont {I.~G.}\ \bibnamefont {Kevrekidis}}, \bibinfo {author} {\bibfnamefont {L.}~\bibnamefont {Lu}}, \bibinfo {author} {\bibfnamefont {P.}~\bibnamefont {Perdikaris}}, \bibinfo {author} {\bibfnamefont {S.}~\bibnamefont {Wang}}, \ and\ \bibinfo {author} {\bibfnamefont {L.}~\bibnamefont {Yang}},\ }\bibfield  {title} {\enquote {\bibinfo {title} {Physics-informed machine learning},}\ }\href {https://doi.org/10.1038/s42254-021-00314-5} {\bibfield  {journal} {\bibinfo  {journal} {Nat. Rev. Phys.}\ }\textbf {\bibinfo {volume} {3}},\ \bibinfo {pages} {422} (\bibinfo {year} {2021})}\BibitemShut {NoStop}%
\bibitem [{\citenamefont {LeCun}\ \emph {et~al.}(2015)\citenamefont {LeCun}, \citenamefont {Bengio},\ and\ \citenamefont {Hinton}}]{lecun2015deep}%
  \BibitemOpen
  \bibfield  {author} {\bibinfo {author} {\bibfnamefont {Y.}~\bibnamefont {LeCun}}, \bibinfo {author} {\bibfnamefont {Y.}~\bibnamefont {Bengio}}, \ and\ \bibinfo {author} {\bibfnamefont {G.}~\bibnamefont {Hinton}},\ }\bibfield  {title} {\enquote {\bibinfo {title} {Deep learning},}\ }\href {https://doi.org/10.1038/nature14539} {\bibfield  {journal} {\bibinfo  {journal} {Nature}\ }\textbf {\bibinfo {volume} {521}},\ \bibinfo {pages} {436} (\bibinfo {year} {2015})}\BibitemShut {NoStop}%
\bibitem [{\citenamefont {Gu}\ \emph {et~al.}(2018)\citenamefont {Gu}, \citenamefont {Wang}, \citenamefont {Kuen}, \citenamefont {Ma}, \citenamefont {Shahroudy}, \citenamefont {Shuai}, \citenamefont {Liu}, \citenamefont {Wang}, \citenamefont {Wang}, \citenamefont {Cai} \emph {et~al.}}]{gu2018recent}%
  \BibitemOpen
  \bibfield  {author} {\bibinfo {author} {\bibfnamefont {J.}~\bibnamefont {Gu}}, \bibinfo {author} {\bibfnamefont {Z.}~\bibnamefont {Wang}}, \bibinfo {author} {\bibfnamefont {J.}~\bibnamefont {Kuen}}, \bibinfo {author} {\bibfnamefont {L.}~\bibnamefont {Ma}}, \bibinfo {author} {\bibfnamefont {A.}~\bibnamefont {Shahroudy}}, \bibinfo {author} {\bibfnamefont {B.}~\bibnamefont {Shuai}}, \bibinfo {author} {\bibfnamefont {T.}~\bibnamefont {Liu}}, \bibinfo {author} {\bibfnamefont {X.}~\bibnamefont {Wang}}, \bibinfo {author} {\bibfnamefont {G.}~\bibnamefont {Wang}}, \bibinfo {author} {\bibfnamefont {J.}~\bibnamefont {Cai}},  \emph {et~al.},\ }\bibfield  {title} {\enquote {\bibinfo {title} {Recent advances in convolutional neural networks},}\ }\href {https://doi.org/10.1016/j.patcog.2017.10.013} {\bibfield  {journal} {\bibinfo  {journal} {Pattern Recognit.}\ }\textbf {\bibinfo {volume} {77}},\ \bibinfo {pages} {354} (\bibinfo {year} {2018})}\BibitemShut {NoStop}%
\bibitem [{\citenamefont {He}\ \emph {et~al.}(2016)\citenamefont {He}, \citenamefont {Zhang}, \citenamefont {Ren},\ and\ \citenamefont {Sun}}]{he2016deep}%
  \BibitemOpen
  \bibfield  {author} {\bibinfo {author} {\bibfnamefont {K.}~\bibnamefont {He}}, \bibinfo {author} {\bibfnamefont {X.}~\bibnamefont {Zhang}}, \bibinfo {author} {\bibfnamefont {S.}~\bibnamefont {Ren}}, \ and\ \bibinfo {author} {\bibfnamefont {J.}~\bibnamefont {Sun}},\ }\bibfield  {title} {\enquote {\bibinfo {title} {Deep residual learning for image recognition},}\ }\href {https://doi.org/10.48550/arXiv.1512.03385} {\bibfield  {journal} {\bibinfo  {journal} {Proc. IEEE Int. Conf. Comput. Vis.(CVPR)}\ ,\ \bibinfo {pages} {770}} (\bibinfo {year} {2016})}\BibitemShut {NoStop}%
\bibitem [{\citenamefont {Shang}\ \emph {et~al.}(2022)\citenamefont {Shang}, \citenamefont {Liu}, \citenamefont {Shao}, \citenamefont {Han}, \citenamefont {Zang}, \citenamefont {Zhang}, \citenamefont {Salama}, \citenamefont {Gao}, \citenamefont {Lee}, \citenamefont {Thomale}, \citenamefont {Manchon}, \citenamefont {Zhang}, \citenamefont {Cui},\ and\ \citenamefont {Schwingenschl\"{o}gl}}]{Shang2022}%
  \BibitemOpen
  \bibfield  {author} {\bibinfo {author} {\bibfnamefont {C.}~\bibnamefont {Shang}}, \bibinfo {author} {\bibfnamefont {S.}~\bibnamefont {Liu}}, \bibinfo {author} {\bibfnamefont {R.}~\bibnamefont {Shao}}, \bibinfo {author} {\bibfnamefont {P.}~\bibnamefont {Han}}, \bibinfo {author} {\bibfnamefont {X.}~\bibnamefont {Zang}}, \bibinfo {author} {\bibfnamefont {X.}~\bibnamefont {Zhang}}, \bibinfo {author} {\bibfnamefont {K.~N.}\ \bibnamefont {Salama}}, \bibinfo {author} {\bibfnamefont {W.}~\bibnamefont {Gao}}, \bibinfo {author} {\bibfnamefont {C.~H.}\ \bibnamefont {Lee}}, \bibinfo {author} {\bibfnamefont {R.}~\bibnamefont {Thomale}}, \bibinfo {author} {\bibfnamefont {A.}~\bibnamefont {Manchon}}, \bibinfo {author} {\bibfnamefont {S.}~\bibnamefont {Zhang}}, \bibinfo {author} {\bibfnamefont {T.~J.}\ \bibnamefont {Cui}}, \ and\ \bibinfo {author} {\bibfnamefont {U.}~\bibnamefont {Schwingenschl\"{o}gl}},\ }\bibfield  {title} {\enquote {\bibinfo {title} {Experimental identification of the second-order non-{H}ermitian skin
  effect with physics-graph-informed machine learning},}\ }\href {https://doi.org/10.1002/advs.202202922} {\bibfield  {journal} {\bibinfo  {journal} {Adv. Sci.}\ }\textbf {\bibinfo {volume} {9}},\ \bibinfo {pages} {2202922} (\bibinfo {year} {2022})}\BibitemShut {NoStop}%
\bibitem [{\citenamefont {Shang}\ \emph {et~al.}(2024)\citenamefont {Shang}, \citenamefont {Liu}, \citenamefont {Jiang}, \citenamefont {Shao}, \citenamefont {Zang}, \citenamefont {Lee}, \citenamefont {Thomale}, \citenamefont {Manchon}, \citenamefont {Cui},\ and\ \citenamefont {Schwingenschl{\"o}gl}}]{shang2024observation}%
  \BibitemOpen
  \bibfield  {author} {\bibinfo {author} {\bibfnamefont {C.}~\bibnamefont {Shang}}, \bibinfo {author} {\bibfnamefont {S.}~\bibnamefont {Liu}}, \bibinfo {author} {\bibfnamefont {C.}~\bibnamefont {Jiang}}, \bibinfo {author} {\bibfnamefont {R.}~\bibnamefont {Shao}}, \bibinfo {author} {\bibfnamefont {X.}~\bibnamefont {Zang}}, \bibinfo {author} {\bibfnamefont {C.~H.}\ \bibnamefont {Lee}}, \bibinfo {author} {\bibfnamefont {R.}~\bibnamefont {Thomale}}, \bibinfo {author} {\bibfnamefont {A.}~\bibnamefont {Manchon}}, \bibinfo {author} {\bibfnamefont {T.~J.}\ \bibnamefont {Cui}}, \ and\ \bibinfo {author} {\bibfnamefont {U.}~\bibnamefont {Schwingenschl{\"o}gl}},\ }\bibfield  {title} {\enquote {\bibinfo {title} {Observation of a higher-order end topological insulator in a real projective lattice},}\ }\href {https://doi.org/10.1002/advs.202303222} {\bibfield  {journal} {\bibinfo  {journal} {Adv. Sci.}\ }\textbf {\bibinfo {volume} {11}},\ \bibinfo {pages} {2303222} (\bibinfo {year} {2024})}\BibitemShut {NoStop}%
\bibitem [{\citenamefont {Bradlyn}\ \emph {et~al.}(2017)\citenamefont {Bradlyn}, \citenamefont {Elcoro}, \citenamefont {Cano}, \citenamefont {Vergniory}, \citenamefont {Wang}, \citenamefont {Felser}, \citenamefont {Aroyo},\ and\ \citenamefont {Bernevig}}]{Bradlyn2017}%
  \BibitemOpen
  \bibfield  {author} {\bibinfo {author} {\bibfnamefont {B.}~\bibnamefont {Bradlyn}}, \bibinfo {author} {\bibfnamefont {L.}~\bibnamefont {Elcoro}}, \bibinfo {author} {\bibfnamefont {J.}~\bibnamefont {Cano}}, \bibinfo {author} {\bibfnamefont {M.~G.}\ \bibnamefont {Vergniory}}, \bibinfo {author} {\bibfnamefont {Z.}~\bibnamefont {Wang}}, \bibinfo {author} {\bibfnamefont {C.}~\bibnamefont {Felser}}, \bibinfo {author} {\bibfnamefont {M.~I.}\ \bibnamefont {Aroyo}}, \ and\ \bibinfo {author} {\bibfnamefont {B.~A.}\ \bibnamefont {Bernevig}},\ }\bibfield  {title} {\enquote {\bibinfo {title} {Topological quantum chemistry},}\ }\href {\doibase 10.1038/nature23268} {\bibfield  {journal} {\bibinfo  {journal} {Nature}\ }\textbf {\bibinfo {volume} {547}},\ \bibinfo {pages} {298} (\bibinfo {year} {2017})}\BibitemShut {NoStop}%
\bibitem [{\citenamefont {Zou}\ \emph {et~al.}(2024)\citenamefont {Zou}, \citenamefont {Chen}, \citenamefont {Meng}, \citenamefont {Ang}, \citenamefont {Zhang},\ and\ \citenamefont {Lee}}]{zou2024experimental}%
  \BibitemOpen
  \bibfield  {author} {\bibinfo {author} {\bibfnamefont {D.}~\bibnamefont {Zou}}, \bibinfo {author} {\bibfnamefont {T.}~\bibnamefont {Chen}}, \bibinfo {author} {\bibfnamefont {H.}~\bibnamefont {Meng}}, \bibinfo {author} {\bibfnamefont {Y.~S.}\ \bibnamefont {Ang}}, \bibinfo {author} {\bibfnamefont {X.}~\bibnamefont {Zhang}}, \ and\ \bibinfo {author} {\bibfnamefont {C.~H.}\ \bibnamefont {Lee}},\ }\bibfield  {title} {\enquote {\bibinfo {title} {Experimental observation of exceptional bound states in a classical circuit network},}\ }\href {https://doi.org/10.1016/j.scib.2024.05.036} {\bibfield  {journal} {\bibinfo  {journal} {Sci. Bull.}\ }\textbf {\bibinfo {volume} {69}},\ \bibinfo {pages} {2194} (\bibinfo {year} {2024})}\BibitemShut {NoStop}%
\bibitem [{\citenamefont {Noh}\ \emph {et~al.}(2018)\citenamefont {Noh}, \citenamefont {Benalcazar}, \citenamefont {Huang}, \citenamefont {Collins}, \citenamefont {Chen}, \citenamefont {Hughes},\ and\ \citenamefont {Rechtsman}}]{noh2018topological}%
  \BibitemOpen
  \bibfield  {author} {\bibinfo {author} {\bibfnamefont {J.}~\bibnamefont {Noh}}, \bibinfo {author} {\bibfnamefont {W.~A.}\ \bibnamefont {Benalcazar}}, \bibinfo {author} {\bibfnamefont {S.}~\bibnamefont {Huang}}, \bibinfo {author} {\bibfnamefont {M.~J.}\ \bibnamefont {Collins}}, \bibinfo {author} {\bibfnamefont {K.~P.}\ \bibnamefont {Chen}}, \bibinfo {author} {\bibfnamefont {T.~L.}\ \bibnamefont {Hughes}}, \ and\ \bibinfo {author} {\bibfnamefont {M.~C.}\ \bibnamefont {Rechtsman}},\ }\bibfield  {title} {\enquote {\bibinfo {title} {Topological protection of photonic mid-gap defect modes},}\ }\href {https://doi.org/10.1038/s41566-018-0179-3} {\bibfield  {journal} {\bibinfo  {journal} {Nat. Photonics}\ }\textbf {\bibinfo {volume} {12}},\ \bibinfo {pages} {408} (\bibinfo {year} {2018})}\BibitemShut {NoStop}%
\bibitem [{\citenamefont {Cao}\ \emph {et~al.}(2018{\natexlab{a}})\citenamefont {Cao}, \citenamefont {Fatemi}, \citenamefont {Fang}, \citenamefont {Watanabe}, \citenamefont {Taniguchi}, \citenamefont {Kaxiras},\ and\ \citenamefont {Jarillo-Herrero}}]{cao2018unconventional}%
  \BibitemOpen
  \bibfield  {author} {\bibinfo {author} {\bibfnamefont {Y.}~\bibnamefont {Cao}}, \bibinfo {author} {\bibfnamefont {V.}~\bibnamefont {Fatemi}}, \bibinfo {author} {\bibfnamefont {S.}~\bibnamefont {Fang}}, \bibinfo {author} {\bibfnamefont {K.}~\bibnamefont {Watanabe}}, \bibinfo {author} {\bibfnamefont {T.}~\bibnamefont {Taniguchi}}, \bibinfo {author} {\bibfnamefont {E.}~\bibnamefont {Kaxiras}}, \ and\ \bibinfo {author} {\bibfnamefont {P.}~\bibnamefont {Jarillo-Herrero}},\ }\bibfield  {title} {\enquote {\bibinfo {title} {Unconventional superconductivity in magic-angle graphene superlattices},}\ }\href {https://doi.org/10.1038/nature26160} {\bibfield  {journal} {\bibinfo  {journal} {Nature}\ }\textbf {\bibinfo {volume} {556}},\ \bibinfo {pages} {43} (\bibinfo {year} {2018}{\natexlab{a}})}\BibitemShut {NoStop}%
\bibitem [{\citenamefont {Cao}\ \emph {et~al.}(2018{\natexlab{b}})\citenamefont {Cao}, \citenamefont {Fatemi}, \citenamefont {Demir}, \citenamefont {Fang}, \citenamefont {Tomarken}, \citenamefont {Luo}, \citenamefont {Sanchez-Yamagishi}, \citenamefont {Watanabe}, \citenamefont {Taniguchi}, \citenamefont {Kaxiras} \emph {et~al.}}]{cao2018correlated}%
  \BibitemOpen
  \bibfield  {author} {\bibinfo {author} {\bibfnamefont {Y.}~\bibnamefont {Cao}}, \bibinfo {author} {\bibfnamefont {V.}~\bibnamefont {Fatemi}}, \bibinfo {author} {\bibfnamefont {A.}~\bibnamefont {Demir}}, \bibinfo {author} {\bibfnamefont {S.}~\bibnamefont {Fang}}, \bibinfo {author} {\bibfnamefont {S.~L.}\ \bibnamefont {Tomarken}}, \bibinfo {author} {\bibfnamefont {J.~Y.}\ \bibnamefont {Luo}}, \bibinfo {author} {\bibfnamefont {J.~D.}\ \bibnamefont {Sanchez-Yamagishi}}, \bibinfo {author} {\bibfnamefont {K.}~\bibnamefont {Watanabe}}, \bibinfo {author} {\bibfnamefont {T.}~\bibnamefont {Taniguchi}}, \bibinfo {author} {\bibfnamefont {E.}~\bibnamefont {Kaxiras}},  \emph {et~al.},\ }\bibfield  {title} {\enquote {\bibinfo {title} {Correlated insulator behaviour at half-filling in magic-angle graphene superlattices},}\ }\href {https://doi.org/10.1038/nature26154} {\bibfield  {journal} {\bibinfo  {journal} {Nature}\ }\textbf {\bibinfo {volume} {556}},\ \bibinfo {pages} {80} (\bibinfo {year}
  {2018}{\natexlab{b}})}\BibitemShut {NoStop}%
\bibitem [{\citenamefont {Lee}\ \emph {et~al.}(2018{\natexlab{b}})\citenamefont {Lee}, \citenamefont {Ho}, \citenamefont {Yang}, \citenamefont {Gong},\ and\ \citenamefont {Papi\ifmmode~\acute{c}\else \'{c}\fi{}}}]{PhysRevLett.121.237401}%
  \BibitemOpen
  \bibfield  {author} {\bibinfo {author} {\bibfnamefont {C.~H.}\ \bibnamefont {Lee}}, \bibinfo {author} {\bibfnamefont {W.~W.}\ \bibnamefont {Ho}}, \bibinfo {author} {\bibfnamefont {B.}~\bibnamefont {Yang}}, \bibinfo {author} {\bibfnamefont {J.}~\bibnamefont {Gong}}, \ and\ \bibinfo {author} {\bibfnamefont {Z.}~\bibnamefont {Papi\ifmmode~\acute{c}\else \'{c}\fi{}}},\ }\bibfield  {title} {\enquote {\bibinfo {title} {Floquet mechanism for non-abelian fractional quantum hall states},}\ }\href {\doibase 10.1103/PhysRevLett.121.237401} {\bibfield  {journal} {\bibinfo  {journal} {Phys. Rev. Lett.}\ }\textbf {\bibinfo {volume} {121}},\ \bibinfo {pages} {237401} (\bibinfo {year} {2018}{\natexlab{b}})}\BibitemShut {NoStop}%
\bibitem [{\citenamefont {Wang}\ \emph {et~al.}(2020)\citenamefont {Wang}, \citenamefont {Zheng}, \citenamefont {Chen}, \citenamefont {Huang}, \citenamefont {Kartashov}, \citenamefont {Torner}, \citenamefont {Konotop},\ and\ \citenamefont {Ye}}]{wang2020localization}%
  \BibitemOpen
  \bibfield  {author} {\bibinfo {author} {\bibfnamefont {P.}~\bibnamefont {Wang}}, \bibinfo {author} {\bibfnamefont {Y.}~\bibnamefont {Zheng}}, \bibinfo {author} {\bibfnamefont {X.}~\bibnamefont {Chen}}, \bibinfo {author} {\bibfnamefont {C.}~\bibnamefont {Huang}}, \bibinfo {author} {\bibfnamefont {Y.~V.}\ \bibnamefont {Kartashov}}, \bibinfo {author} {\bibfnamefont {L.}~\bibnamefont {Torner}}, \bibinfo {author} {\bibfnamefont {V.~V.}\ \bibnamefont {Konotop}}, \ and\ \bibinfo {author} {\bibfnamefont {F.}~\bibnamefont {Ye}},\ }\bibfield  {title} {\enquote {\bibinfo {title} {Localization and delocalization of light in photonic moir{\'e} lattices},}\ }\href {https://doi.org/10.1038/s41586-019-1851-6} {\bibfield  {journal} {\bibinfo  {journal} {Nature}\ }\textbf {\bibinfo {volume} {577}},\ \bibinfo {pages} {42} (\bibinfo {year} {2020})}\BibitemShut {NoStop}%
\bibitem [{\citenamefont {Cai}\ and\ \citenamefont {Wang}(2020)}]{Cai2020}%
  \BibitemOpen
  \bibfield  {author} {\bibinfo {author} {\bibfnamefont {H.}~\bibnamefont {Cai}}\ and\ \bibinfo {author} {\bibfnamefont {D.}~\bibnamefont {Wang}},\ }\bibfield  {title} {\enquote {\bibinfo {title} {Topological phases of quantized light},}\ }\href {http://dx.doi.org/10.1093/nsr/nwaa196} {\bibfield  {journal} {\bibinfo  {journal} {Natl. Sci. Rev.}\ }\textbf {\bibinfo {volume} {8}},\ \bibinfo {pages} {nwaa196} (\bibinfo {year} {2020})}\BibitemShut {NoStop}%
\bibitem [{\citenamefont {Yang}\ \emph {et~al.}(2024{\natexlab{b}})\citenamefont {Yang}, \citenamefont {Li}, \citenamefont {Yang}, \citenamefont {Xie}, \citenamefont {Zhang}, \citenamefont {Yuan}, \citenamefont {Cai}, \citenamefont {Wang},\ and\ \citenamefont {Gao}}]{yang2024realization}%
  \BibitemOpen
  \bibfield  {author} {\bibinfo {author} {\bibfnamefont {J.}~\bibnamefont {Yang}}, \bibinfo {author} {\bibfnamefont {Y.}~\bibnamefont {Li}}, \bibinfo {author} {\bibfnamefont {Y.}~\bibnamefont {Yang}}, \bibinfo {author} {\bibfnamefont {X.}~\bibnamefont {Xie}}, \bibinfo {author} {\bibfnamefont {Z.}~\bibnamefont {Zhang}}, \bibinfo {author} {\bibfnamefont {J.}~\bibnamefont {Yuan}}, \bibinfo {author} {\bibfnamefont {H.}~\bibnamefont {Cai}}, \bibinfo {author} {\bibfnamefont {D.-W.}\ \bibnamefont {Wang}}, \ and\ \bibinfo {author} {\bibfnamefont {F.}~\bibnamefont {Gao}},\ }\bibfield  {title} {\enquote {\bibinfo {title} {Realization of all-band-flat photonic lattices},}\ }\href {https://doi.org/10.1038/s41467-024-45580-w} {\bibfield  {journal} {\bibinfo  {journal} {Nat. Commun.}\ }\textbf {\bibinfo {volume} {15}},\ \bibinfo {pages} {1484} (\bibinfo {year} {2024}{\natexlab{b}})}\BibitemShut {NoStop}%
\bibitem [{\citenamefont {Deng}\ \emph {et~al.}(2022)\citenamefont {Deng}, \citenamefont {Dong}, \citenamefont {Zhang}, \citenamefont {Wu}, \citenamefont {Yuan}, \citenamefont {Zhu}, \citenamefont {Jin}, \citenamefont {Li}, \citenamefont {Wang}, \citenamefont {Cai} \emph {et~al.}}]{deng2022observing}%
  \BibitemOpen
  \bibfield  {author} {\bibinfo {author} {\bibfnamefont {J.}~\bibnamefont {Deng}}, \bibinfo {author} {\bibfnamefont {H.}~\bibnamefont {Dong}}, \bibinfo {author} {\bibfnamefont {C.}~\bibnamefont {Zhang}}, \bibinfo {author} {\bibfnamefont {Y.}~\bibnamefont {Wu}}, \bibinfo {author} {\bibfnamefont {J.}~\bibnamefont {Yuan}}, \bibinfo {author} {\bibfnamefont {X.}~\bibnamefont {Zhu}}, \bibinfo {author} {\bibfnamefont {F.}~\bibnamefont {Jin}}, \bibinfo {author} {\bibfnamefont {H.}~\bibnamefont {Li}}, \bibinfo {author} {\bibfnamefont {Z.}~\bibnamefont {Wang}}, \bibinfo {author} {\bibfnamefont {H.}~\bibnamefont {Cai}},  \emph {et~al.},\ }\bibfield  {title} {\enquote {\bibinfo {title} {Observing the quantum topology of light},}\ }\href {https://doi.org/10.1126/science.ade6219} {\bibfield  {journal} {\bibinfo  {journal} {Science}\ }\textbf {\bibinfo {volume} {378}},\ \bibinfo {pages} {966} (\bibinfo {year} {2022})}\BibitemShut {NoStop}%
\bibitem [{\citenamefont {Olekhno}\ \emph {et~al.}(2020)\citenamefont {Olekhno}, \citenamefont {Kretov}, \citenamefont {Stepanenko}, \citenamefont {Ivanova}, \citenamefont {Yaroshenko}, \citenamefont {Puhtina}, \citenamefont {Filonov}, \citenamefont {Cappello}, \citenamefont {Matekovits},\ and\ \citenamefont {Gorlach}}]{Olekhno2020Topolectrical}%
  \BibitemOpen
  \bibfield  {author} {\bibinfo {author} {\bibfnamefont {N.~A.}\ \bibnamefont {Olekhno}}, \bibinfo {author} {\bibfnamefont {E.~I.}\ \bibnamefont {Kretov}}, \bibinfo {author} {\bibfnamefont {A.~A.}\ \bibnamefont {Stepanenko}}, \bibinfo {author} {\bibfnamefont {P.~A.}\ \bibnamefont {Ivanova}}, \bibinfo {author} {\bibfnamefont {V.~V.}\ \bibnamefont {Yaroshenko}}, \bibinfo {author} {\bibfnamefont {E.~M.}\ \bibnamefont {Puhtina}}, \bibinfo {author} {\bibfnamefont {D.~S.}\ \bibnamefont {Filonov}}, \bibinfo {author} {\bibfnamefont {B.}~\bibnamefont {Cappello}}, \bibinfo {author} {\bibfnamefont {L.}~\bibnamefont {Matekovits}}, \ and\ \bibinfo {author} {\bibfnamefont {M.~A.}\ \bibnamefont {Gorlach}},\ }\bibfield  {title} {\enquote {\bibinfo {title} {Topological edge states of interacting photon pairs emulated in a topolectrical circuit},}\ }\href {\doibase 10.1038/s41467-020-14994-7} {\bibfield  {journal} {\bibinfo  {journal} {Nature Communications}\ }\textbf {\bibinfo {volume} {11}},\ \bibinfo {pages} {1436}
  (\bibinfo {year} {2020})}\BibitemShut {NoStop}%
\bibitem [{\citenamefont {Zhou}\ \emph {et~al.}(2023)\citenamefont {Zhou}, \citenamefont {Zhang}, \citenamefont {Sun},\ and\ \citenamefont {Zhang}}]{zhou2023observation}%
  \BibitemOpen
  \bibfield  {author} {\bibinfo {author} {\bibfnamefont {X.}~\bibnamefont {Zhou}}, \bibinfo {author} {\bibfnamefont {W.}~\bibnamefont {Zhang}}, \bibinfo {author} {\bibfnamefont {H.}~\bibnamefont {Sun}}, \ and\ \bibinfo {author} {\bibfnamefont {X.}~\bibnamefont {Zhang}},\ }\bibfield  {title} {\enquote {\bibinfo {title} {Observation of flat-band localization and topological edge states induced by effective strong interactions in electrical circuit networks},}\ }\href {\doibase 10.1103/PhysRevB.107.035152} {\bibfield  {journal} {\bibinfo  {journal} {Physical Review B}\ }\textbf {\bibinfo {volume} {107}},\ \bibinfo {pages} {035152} (\bibinfo {year} {2023})}\BibitemShut {NoStop}%
\bibitem [{\citenamefont {Anderson}(1958)}]{physRev.109.1492}%
  \BibitemOpen
  \bibfield  {author} {\bibinfo {author} {\bibfnamefont {P.~W.}\ \bibnamefont {Anderson}},\ }\bibfield  {title} {\enquote {\bibinfo {title} {Absence of diffusion in certain random lattices},}\ }\href {\doibase 10.1103/PhysRev.109.1492} {\bibfield  {journal} {\bibinfo  {journal} {Phys. Rev.}\ }\textbf {\bibinfo {volume} {109}},\ \bibinfo {pages} {1492} (\bibinfo {year} {1958})}\BibitemShut {NoStop}%
\bibitem [{\citenamefont {Kondov}\ \emph {et~al.}(2011)\citenamefont {Kondov}, \citenamefont {McGehee}, \citenamefont {Zirbel},\ and\ \citenamefont {DeMarco}}]{kondov2011three}%
  \BibitemOpen
  \bibfield  {author} {\bibinfo {author} {\bibfnamefont {S.}~\bibnamefont {Kondov}}, \bibinfo {author} {\bibfnamefont {W.}~\bibnamefont {McGehee}}, \bibinfo {author} {\bibfnamefont {J.}~\bibnamefont {Zirbel}}, \ and\ \bibinfo {author} {\bibfnamefont {B.}~\bibnamefont {DeMarco}},\ }\bibfield  {title} {\enquote {\bibinfo {title} {Three-dimensional anderson localization of ultracold matter},}\ }\href {https://doi.org/10.1126/science.1209019} {\bibfield  {journal} {\bibinfo  {journal} {Science}\ }\textbf {\bibinfo {volume} {334}},\ \bibinfo {pages} {66} (\bibinfo {year} {2011})}\BibitemShut {NoStop}%
\bibitem [{\citenamefont {Yu}\ \emph {et~al.}(2021)\citenamefont {Yu}, \citenamefont {Qiu}, \citenamefont {Chong}, \citenamefont {Torquato},\ and\ \citenamefont {Park}}]{yu2021engineered}%
  \BibitemOpen
  \bibfield  {author} {\bibinfo {author} {\bibfnamefont {S.}~\bibnamefont {Yu}}, \bibinfo {author} {\bibfnamefont {C.-W.}\ \bibnamefont {Qiu}}, \bibinfo {author} {\bibfnamefont {Y.}~\bibnamefont {Chong}}, \bibinfo {author} {\bibfnamefont {S.}~\bibnamefont {Torquato}}, \ and\ \bibinfo {author} {\bibfnamefont {N.}~\bibnamefont {Park}},\ }\bibfield  {title} {\enquote {\bibinfo {title} {Engineered disorder in photonics},}\ }\href {https://doi.org/10.1038/s41578-020-00263-y} {\bibfield  {journal} {\bibinfo  {journal} {Nat. Rev. Mat.}\ }\textbf {\bibinfo {volume} {6}},\ \bibinfo {pages} {226} (\bibinfo {year} {2021})}\BibitemShut {NoStop}%
\bibitem [{\citenamefont {Ho}\ \emph {et~al.}(2020)\citenamefont {Ho}, \citenamefont {Jain},\ and\ \citenamefont {Abbeel}}]{ho2020denoising}%
  \BibitemOpen
  \bibfield  {author} {\bibinfo {author} {\bibfnamefont {J.}~\bibnamefont {Ho}}, \bibinfo {author} {\bibfnamefont {A.}~\bibnamefont {Jain}}, \ and\ \bibinfo {author} {\bibfnamefont {P.}~\bibnamefont {Abbeel}},\ }\bibfield  {title} {\enquote {\bibinfo {title} {Denoising diffusion probabilistic models},}\ }\href {https://dl.acm.org/doi/abs/10.5555/3495724.3496298} {\bibfield  {journal} {\bibinfo  {journal} {Adv. Neural Inf. Process. Syst.}\ }\textbf {\bibinfo {volume} {33}},\ \bibinfo {pages} {6840} (\bibinfo {year} {2020})}\BibitemShut {NoStop}%
\bibitem [{\citenamefont {Santini}\ \emph {et~al.}(2020)\citenamefont {Santini}, \citenamefont {Battaglioni}, \citenamefont {Baldi},\ and\ \citenamefont {Chiaraluce}}]{santini2020analysis}%
  \BibitemOpen
  \bibfield  {author} {\bibinfo {author} {\bibfnamefont {P.}~\bibnamefont {Santini}}, \bibinfo {author} {\bibfnamefont {M.}~\bibnamefont {Battaglioni}}, \bibinfo {author} {\bibfnamefont {M.}~\bibnamefont {Baldi}}, \ and\ \bibinfo {author} {\bibfnamefont {F.}~\bibnamefont {Chiaraluce}},\ }\bibfield  {title} {\enquote {\bibinfo {title} {Analysis of the error correction capability of ldpc and mdpc codes under parallel bit-flipping decoding and application to cryptography},}\ }\href {https://doi.org/10.1109/TCOMM.2020.2987898} {\bibfield  {journal} {\bibinfo  {journal} {IEEE Trans. Commun.}\ }\textbf {\bibinfo {volume} {68}},\ \bibinfo {pages} {4648} (\bibinfo {year} {2020})}\BibitemShut {NoStop}%
\bibitem [{\citenamefont {Sohn}\ \emph {et~al.}(2015)\citenamefont {Sohn}, \citenamefont {Lee},\ and\ \citenamefont {Yan}}]{sohn2015learning}%
  \BibitemOpen
  \bibfield  {author} {\bibinfo {author} {\bibfnamefont {K.}~\bibnamefont {Sohn}}, \bibinfo {author} {\bibfnamefont {H.}~\bibnamefont {Lee}}, \ and\ \bibinfo {author} {\bibfnamefont {X.}~\bibnamefont {Yan}},\ }\bibfield  {title} {\enquote {\bibinfo {title} {Learning structured output representation using deep conditional generative models},}\ }\href {https://dl.acm.org/doi/10.5555/2969442.2969628} {\bibfield  {journal} {\bibinfo  {journal} {Adv. Neural Inf. Process. Syst.}\ }\textbf {\bibinfo {volume} {2}},\ \bibinfo {pages} {3483} (\bibinfo {year} {2015})}\BibitemShut {NoStop}%
\bibitem [{\citenamefont {Gao}\ \emph {et~al.}(2020)\citenamefont {Gao}, \citenamefont {Al-Sarawi},\ and\ \citenamefont {Abbott}}]{gao2020physical}%
  \BibitemOpen
  \bibfield  {author} {\bibinfo {author} {\bibfnamefont {Y.}~\bibnamefont {Gao}}, \bibinfo {author} {\bibfnamefont {S.~F.}\ \bibnamefont {Al-Sarawi}}, \ and\ \bibinfo {author} {\bibfnamefont {D.}~\bibnamefont {Abbott}},\ }\bibfield  {title} {\enquote {\bibinfo {title} {Physical unclonable functions},}\ }\href {https://doi.org/10.1038/s41928-020-0372-5} {\bibfield  {journal} {\bibinfo  {journal} {Nat. Electron.}\ }\textbf {\bibinfo {volume} {3}},\ \bibinfo {pages} {81} (\bibinfo {year} {2020})}\BibitemShut {NoStop}%
\bibitem [{\citenamefont {Shih}(2017)}]{shih2017digital}%
  \BibitemOpen
  \bibfield  {author} {\bibinfo {author} {\bibfnamefont {F.~Y.}\ \bibnamefont {Shih}},\ }\href {https://doi.org/10.1201/9781315121109} {\emph {\bibinfo {title} {Digital watermarking and steganography: fundamentals and techniques}}}\ (\bibinfo  {publisher} {CRC press},\ \bibinfo {address} {Boca Raton},\ \bibinfo {year} {2017})\BibitemShut {NoStop}%
\bibitem [{\citenamefont {Shamsoshoara}\ \emph {et~al.}(2020)\citenamefont {Shamsoshoara}, \citenamefont {Korenda}, \citenamefont {Afghah},\ and\ \citenamefont {Zeadally}}]{shamsoshoara2020survey}%
  \BibitemOpen
  \bibfield  {author} {\bibinfo {author} {\bibfnamefont {A.}~\bibnamefont {Shamsoshoara}}, \bibinfo {author} {\bibfnamefont {A.}~\bibnamefont {Korenda}}, \bibinfo {author} {\bibfnamefont {F.}~\bibnamefont {Afghah}}, \ and\ \bibinfo {author} {\bibfnamefont {S.}~\bibnamefont {Zeadally}},\ }\bibfield  {title} {\enquote {\bibinfo {title} {A survey on physical unclonable function (puf)-based security solutions for internet of things},}\ }\href {https://doi.org/10.1016/j.comnet.2020.107593} {\bibfield  {journal} {\bibinfo  {journal} {Comput. Netw.}\ }\textbf {\bibinfo {volume} {183}},\ \bibinfo {pages} {107593} (\bibinfo {year} {2020})}\BibitemShut {NoStop}%
\bibitem [{\citenamefont {Shamir}(1979)}]{shamir1979share}%
  \BibitemOpen
  \bibfield  {author} {\bibinfo {author} {\bibfnamefont {A.}~\bibnamefont {Shamir}},\ }\bibfield  {title} {\enquote {\bibinfo {title} {How to share a secret},}\ }\href {https://dl.acm.org/doi/abs/10.1145/359168.359176} {\bibfield  {journal} {\bibinfo  {journal} {Commun. the ACM}\ }\textbf {\bibinfo {volume} {22}},\ \bibinfo {pages} {612} (\bibinfo {year} {1979})}\BibitemShut {NoStop}%
\bibitem [{\citenamefont {Wicker}\ and\ \citenamefont {Bhargava}(1999)}]{wicker1999reed}%
  \BibitemOpen
  \bibfield  {author} {\bibinfo {author} {\bibfnamefont {S.~B.}\ \bibnamefont {Wicker}}\ and\ \bibinfo {author} {\bibfnamefont {V.~K.}\ \bibnamefont {Bhargava}},\ }\href {https://dl.acm.org/doi/10.5555/554634} {\emph {\bibinfo {title} {Reed-Solomon codes and their applications}}}\ (\bibinfo  {publisher} {John Wiley \& Sons},\ \bibinfo {year} {1999})\BibitemShut {NoStop}%
\end{thebibliography}
\end{document}